\documentclass[%
 a4paper, twocolumn,11pt
amsmath,amssymb
aps,floatfix,superscriptaddress,
]
{revtex4-2}
\pdfoutput = 1

\usepackage[utf8]{inputenc}
\usepackage[english]{babel}
\usepackage[T1]{fontenc}
\usepackage{xkeyval}
\usepackage{graphicx}%
\usepackage[utf8]{inputenc}
\usepackage{tikz}
\usepackage{tikz-cd}
\usepackage{pgfplots}
\pgfplotsset{compat=newest}
\usepackage{lmodern}
\usepackage[export]{adjustbox}%
\usepackage{soul}
\usepackage{dcolumn}%
\usepackage{bm}
\usepackage{epstopdf}
 \usepackage[usenames,dvipsnames]{pstricks}
 \usepackage{epsfig}
 \usepackage{pst-grad} %
 \usepackage{pst-plot} %
 \usepackage[space]{grffile} %
 \usepackage{etoolbox} %
 \makeatletter %
 \patchcmd\Gread@eps{\@inputcheck#1 }{\@inputcheck"#1"\relax}{}{}
 \makeatother
\usepackage{booktabs}
\usepackage{boxhandler}
\usepackage{amsmath}
\usepackage{subcaption}
\usepackage{braket}
\usepackage{xcolor}
\usepackage{float}
\usepackage[normalem]{ulem}

\makeatletter
\let\newfloat\newfloat@ltx
\makeatother
\usepackage{algpseudocode}
\usepackage{algorithm}
\usepackage{amsfonts}
\usepackage{hyperref}
\hypersetup{
  colorlinks   = true,    %
  urlcolor     = blue,    %
  linkcolor    = blue,    %
  citecolor    = blue      %
}
\newcommand{\algorithmicbreak}{\textbf{break}}

\usepackage{multirow}
\usepackage{enumitem}

\newtheorem{definition}{Definition}

\begin{document}

\title{An almost-linear time decoding algorithm for quantum\\ LDPC codes under circuit-level noise}

\author{Antonio {deMarti iOlius}}
\email{toni.demarti@gmail.com}
\affiliation{Department of Basic Sciences, Tecnun - University of Navarra, 20018 San Sebastian, Spain.}
\author{Imanol {Etxezarreta Martinez}}
\email{ietxezarretam@gmail.com}
\affiliation{Independent researcher, San Sebastian, Spain.}
\author{Joschka Roffe}
\email{joschka@roffe.eu}
\affiliation{Quantum Software Lab, University of Edinburgh, United Kingdom.}
\author{Josu {Etxezarreta Martinez}}
\email{jetxezarreta@unav.es}
\affiliation{Department of Basic Sciences, Tecnun - University of Navarra, 20018 San Sebastian, Spain.}

\begin{abstract}
    Fault-tolerant quantum computers must be designed in conjunction with classical co-processors that decode quantum error correction measurement information in real-time. In this work, we introduce the belief propagation plus ordered Tanner forest (BP+OTF) algorithm as an almost-linear time decoder for quantum low-density parity-check codes. The OTF post-processing stage removes qubits from the decoding graph until it has a tree-like structure. Provided that the resultant loop-free OTF graph supports a subset of qubits that can generate the syndrome, BP decoding is then guaranteed to converge. To enhance performance under circuit-level noise, we introduce a technique for sparsifying detector error models. This method uses a transfer matrix to map soft information from the full detector graph to the sparsified graph, preserving critical error propagation information from the syndrome extraction circuit. Our BP+OTF implementation first applies standard BP to the full detector graph, followed by BP+OTF post-processing on the sparsified graph. Numerical simulations show that the BP+OTF decoder achieves similar logical error suppression compared to state-of-the-art inversion-based and matching decoders for bivariate bicycle and surface codes, respectively, while maintaining almost-linear runtime complexity across all stages.

\end{abstract}

\keywords{Quantum error correction, qldpc codes, decoherence}
\maketitle
\section{Introduction}

Quantum error correction (QEC) is the key to operationally useful quantum computing and will enable the construction of fault-tolerant systems with capabilities well in excess of classical computing technologies \cite{QECbeginners}. The past decade has seen significant progress in experimental realisations of QEC, recently highlighted by the first demonstration by Google Quantum AI of a surface code logical qubit operating below the break-even point \cite{googleSurface24}. However, significant barriers remain to the practical realisation of QEC at scale. A particularly pertinent challenge relates to the problem of real-time decoding: the operation of a QEC code relies upon the ability to process vast quantities of syndrome information at timescales as low as $1\mu s$ \cite{surfacecodeETH} using classical co-processors \cite{decoders}. As such, fast and accurate decoding algorithms are a critical component of the quantum computing stack.

Surface code QEC architectures are conceptually simple and can be implemented using local interactions between qubits arranged in a grid. However, a disadvantage of surface codes is that they have poor encoding rate and will require large qubit counts to achieve logical error rates in the \textit{teraquop} regime (logical failure rate $p_L<10^{-12}$) that is thought to be necessary to run algorithms that offer quantum advantage. Quantum low-density parity-check (QLDPC) codes are an alternative QEC protocol to the surface code and are suitable for quantum computing technologies that have high qubit connectivity. Numerical simulations of QLDPC code architectures suggest that they will require approximately $10\times$ fewer qubits compared to surface code architectures \cite{BBcodes,xu2023constantoverheadfaulttolerantquantumcomputation,scruby2024highthresholdlowoverheadsingleshotdecodable}. However, the design of efficient decoding algorithms for QLDPC codes remains an open problem and a bottleneck to their implementation in experiment.

The belief propagation (BP) decoder is ubiquitous in classical communications, enabling real-time decoding for LDPC protocols underpinning technologies such as 5G, WiFi, and Ethernet ~\cite{richardson_design_2018}. The primary advantage of BP decoders lies in their speed: they can be executed on distributed hardware and have a linear worst-case runtime complexity of $\mathcal{O}(n)$ with respect to the code block length $n$ \cite{mackay1997near}. However, when applied to QLDPC codes, BP decoders face significant challenges. These challenges stem from the presence of degenerate quantum errors and the emergence of trapping sets, which can prevent the BP decoder from converging \cite{degen,Raveendran2021trappingsetsof}. To address these issues, decoders for QLDPC codes typically augment BP with post-processing routines that are invoked when the main algorithm fails \cite{Grospellier_2021,panteleev_bposd,Joschka_decoding_across,Scruby_2023,AmbiguityClusteringOSD,closedbranch,joschkaLSD,bpgdg,DTD,tesseract,koutsioumpas2025automorphismensembledecodingquantum,relayBP,Maan2025}. The most accurate of these two-stage algorithms — ordered statistics decoding (BP+OSD) \cite{panteleev_bposd,Joschka_decoding_across}, ambiguity clustering (BP+AC) \cite{AmbiguityClusteringOSD}, and localized statistics decoding (BP+LSD) \cite{joschkaLSD} — employ matrix inversion techniques to solve the decoding problem during post-processing. While these inversion-based decoders achieve high error thresholds, they also increase the worst-case runtime complexity to $\mathcal{O}(n^3)$. This is significantly slower than the $\mathcal{O}(n)$ runtime of standard BP.

In this work, we present the ordered Tanner forest (OTF) algorithm as a post-processor to belief propagation that does not require matrix inversion. The core idea behind OTF post-processing is based on the fact that the BP algorithm is guaranteed to converge when applied to decoding graphs with a tree-like structure. The OTF post-processor leverages this by constructing an ordered spanning tree of the QEC code's decoding (Tanner) graph, giving priority to nodes that were assigned a high probability of supporting an error during the initial BP decoding attempt. This spanning tree can contain multiple disconnected components, so we refer to it as an \textit{ordered Tanner forest}. Once the ordered Tanner forest is formed, the BP algorithm can be applied directly to it to identify a decoding solution that aligns with the measured syndrome.

The OTF post-processor employs a modified version of Kruskal's algorithm to find the ordered Tanner forest, with a worst-case runtime complexity of $\mathcal{O}(n\log(n))$ \cite{kruskal1956shortest}. This process involves searching through the nodes of the decoding graph and eliminating those that would introduce loops. A second round of BP is then run on the ordered Tanner forest, with a linear time complexity proportional to the number of remaining nodes. As a result, the combined BP+OTF decoder achieves an almost-linear runtime relative to the code's block length.

In practice, QEC decoding algorithms are run on a \textit{detector} graph  or detector error model which relates error mechanism in the QEC circuit to the measured syndromes \cite{detectorsPaper}. A problem with applying OTF post-processing directly to the detector graph is that the columns are typically high-weight compared to the \textit{code capacity} or \textit{phenomenological} graph for the same code. As a consequence, it is often the case that many graph nodes need to be discarded in the search for the ordered Tanner forest, sometimes to the point where the remaining graph no-longer contains enough qubits to support a solution to the syndrome. To address this problem, we propose a novel procedure for mapping a detector graph to a \textit{sparsified} detector graph with fewer short-length loops. When applied to this sparsified detector graph, OTF post-processing is more likely to succeed.

We propose a three-stage decoder specifically optimized for  circuit-level noise which we refer to as BP+BP+OTF. First, standard BP decoding is applied to the full detector graph. Second, the soft-information output of the first BP round is mapped to the sparsified detector graph via a pre-defined transfer matrix. Using this soft-information as prior-information, BP decoding is applied a second time to the sparsified detector graph. Third, OTF post-processing is run on the sparsified detector graph guided by the soft-information output of the previous round of BP. Note that the BP+BP+OTF decoder terminates at the first successful decoding. E.g., the final round of OTF will not be invoked if either of the preceding rounds of BP succeeds.

To benchmark the BP+BP+OTF decoder, we run Monte Carlo simulations under depolarising circuit-level noise for bivariate bicycle codes \cite{BBcodes} as well as for surface codes. Our results demonstrate that the BP+BP+OTF decoder achieves comparable error suppression to state-of-the-art decoders such as BP+OSD and minimum-weight perfect-matching. Notably, our implementation runs more than $10\times$ faster compared to the BP+OSD implementation from \cite{joschkaLDPC} when decoding bivariate bicycle codes under certain noise regimes. Furthermore, our results show that BP+BP+OTF can reliably decode surface codes using a very small number of iterations of the minimum-sum algorithm.

One of the key advantages of BP+BP+OTF is its simplicity: the decoder requires just three applications of a standard BP decoder and a single application of a modification of the well-known Kruskal algorithm for generating the ordered spanning forest. This straightforward design allows for the construction of highly efficient hardware implementations using off-the-shelf application-specific integrated circuits (ASICs). As a result, the BP+OTF decoder is an appealing option for real-time decoding.

\section{Preliminaries}

\subsection{Calderbank-Shor-Steane codes}

Calderbank-Shor-Steane (CSS) codes describe a large family of quantum error correction protocols that can be expressed in terms of a pair of classical binary codes, $\mathcal{Q(H_X,H_Z)}$. The $H_X$ matrix describes $X$-type stabilisers that detect phase-flip errors and the $H_Z$ matrix $Z$-type stabilisers that detect bit-flip errors. If $H_{X/Z}$ are sparse, we can consider the CSS code to be a quantum LDPC code.

The decoding task for CSS codes involves solving two syndrome equations for $X/Z$ errors
\begin{equation}
    H_{X/Z}\cdot \textbf{e}_{Z/X} \mod 2= \textbf{s}_{X/Z}\rm.
\end{equation}
As the above syndrome equations amount to a pair of classical decoding problems, it is possible to use existing classical decoders to directly decode CSS codes. Note that the above syndrome equation is over binary field modulo-$2$. From this point onwards, we will assume all linear algebra operations are performed modulo-$2$ unless stated otherwise.

\subsection{Belief propagation decoding}

Belief propagation (BP) is common decoding algorithm in classical communication technologies \cite{mackay1997near,richardson_design_2018}. Decoders based on BP exploit the sparsity of the LDPC code's parity check matrix to efficiently solve the bit-wise decoding problem
\begin{equation}
    p(\mathbf{e}_i) = \sum_{\sim i} p(\mathbf{e}|\mathbf{s})\rm.
\end{equation}
The specific advantage of the BP decoder lies in its speed. The message passing operations that underpin the algorithm occur locally between the nodes of a graphical representation of the parity check matrix known as a Tanner graph, and can be run in parallel on distributed hardware. In this setting, BP has a worst-case runtime complexity $\mathcal{O}(n)$ in the code's block-length $n$.

The product-sum algorithm is a variant of the BP decoder that is known to yield exact marginals for parity check matrices that have tree-like Tanner graphs. In practice, however, most high-rate LDPC codes contain loops that compromise the performance of product-sum BP and can prevent it from converging to a valid solution. A challenge in designing LDPC codes lies in finding parity check matrices that are sufficiently loop-free and sparse.

For quantum LDPC codes, it is particularly difficult to design $H_X$ and $H_Z$ matrices that are a well-suited for BP decoding. The reason for this can be attributed to degenerate errors that lead to cyclic dependencies in the decoding graph. As such, BP-based algorithms typically do not work as effective decoders for quantum LDPC codes. Indeed, a standard implementation of product-sum BP fails to yield a threshold for the surface code \cite{Joschka_decoding_across}. In practice, quantum decoders require BP to be augmented with a post-processing routine to achieve satisfactory performance in terms of both threshold and sub-threshold error suppression \cite{Grospellier_2021,panteleev_bposd,Joschka_decoding_across,Scruby_2023,AmbiguityClusteringOSD,closedbranch,joschkaLSD}.

\subsection{Decoding circuit-level noise and the detector error model} \label{sec:circuit-level-decoding}

In classical error correction, the decoding problem involves directly solving the syndrome equation $H\cdot \textbf{e} = \textbf{s}$, where $H$ is the parity-check matrix of the code. In quantum error correction, however, an additional layer of complexity arises due to the fact that syndromes are measured using noisy circuits, leading to error propagation that is not described by the CSS code's $H_{X/Z}$ matrices. Instead, the circuit-level decoding problem is characterised by a binary matrix, $H_{CL}$, where the columns correspond to error mechanisms within the circuit, and the rows correspond to syndrome measurements. For instance, a non-zero entry at position $(H_{CL})_{ij}$ indicates that error mechanism $j$ triggers syndrome measurement $i$. Once the circuit-level matrix $H_{CL}$ is constructed, the decoding problem becomes equivalent to that of decoding a classical linear code, specifically solving $H_{CL}\cdot \textbf{e} = \textbf{s}$. The key conceptual difference is that the columns of $H_{CL}$ represent circuit error mechanisms, rather than Pauli-$X/Z$ errors, as is the case in code capacity decoding using the $H_{X/Z}$ CSS matrices.

In practice, QEC protocols operate by repeatedly measuring the same stabiliser extraction circuit over time. This repeating structure can be leveraged to simplify the circuit-level decoding problem by mapping it to a \textit{detector error model} \cite{detectorsPaper}. A detector vector is defined as a linear combination of syndrome measurements that sums to zero when there are no errors in the circuit. In a repeated stabiliser measurement circuit, the most intuitive choice of detectors involves comparing the parity between consecutive syndrome measurements. For example, if the same check yields a value of $1$ in two consecutive rounds, the detector value will be trivial. The resulting detector vector $\textbf{s}_D$ is therefore sparser than the original syndrome $\textbf{s}$. Correspondingly, the circuit-level decoding matrix $H_{CL}$ is replaced with a detector matrix $H_{dem}$, as a $d\times n$ binary matrix, where $d$ is the number of measurements (detector elements) and $n$ is the number of possible errors in the circuit \footnote{Strictly, the number of columns $n$ is not exactly the number of possible circuit errors since errors that have the same detector vectors are merged in single columns, adding their a priori probabilities.}. Here, the rows correspond to detector measurements rather than syndromes. Typically, $H_{dem}$ is also sparser than $H_{CL}$. This sparser structure of the detector error model, described by the equation $H_{dem} \cdot \textbf{e} = \textbf{s}_D$, makes the decoding problem more amenable to BP decoding. From this point onwards, we will refer to $\mathbf{s}$ as a syndrome (code capacity, phenomenological noise) or a detector (circuit-level noise) interchangeably.

Once a set of fault-mechanisms $\mathbf{\hat{e}}$ have been identified by decoding the detector error model $H_{dem} \cdot \mathbf{e} = \textbf{s}$, the task remains to determine their logical action. This can be analysed using the logical observable matrix, $O_{dem} \in \mathbb{F}^{k \times n}$, where $k$ is the number of logical qubits encoded, and $n$ represents the number of possible fault mechanisms in the detector error model. This matrix links errors in the circuit to measured logical observables, which are defined as binary sums of measurements that correspond to the outcomes of logical operator measurements for any logical state encoded by the QEC code. The logical observable matrix associated with the detector error model is crucial for verifying whether the error estimate produced by the decoder successfully corrects the actual faults that occurred. A logical error is detected if the logical observable derived from the decoded error does not match the true logical observable, i.e., if $O_{dem} \cdot \mathbf{e}\neq O_{dem} \cdot \mathbf{\hat{e}}$.

The detector error model, as well as the the logical observable matrix for a specific detector error model related to the syndrome extraction circuit can be generated using a stabiliser simulator \cite{stim,Fang_2024}.

\section{The Ordered Tanner Forest decoder, an almost-linear time post-processor}

In this section, we introduce the ordered Tanner forest (OTF) decoder as an almost-linear post-processor for general QLDPC codes. The OTF decoder is a post-processing algorithm that is called if the original round of BP decoding fails, i.e., when the estimated BP error $\mathbf{\hat{e}}_{BP}$ does not satisfy the syndrome equation, $H\cdot\mathbf{\hat{e}}_{BP}\neq \textbf{s}$. Being a post-processor, OTF requires the posterior probabilities coming from a BP decoding stage (or some process able to provide soft reliability information) and is only invoked when BP does not converge. Based on the received soft information, we sort the columns of the parity check matrix from the least reliable to the most reliable \cite{decoders}. The OTF post-processor then works by deleting columns in the ordered parity check matrix to obtain a reduced parity check matrix, $H_{otf}$, that maps to a Tanner graph that is completely loop free. Note that the resultant spanning tree may contain multiple disconnected trees, so we refer to the returned structure as an \textit{ordered Tanner forest} (OTF). A product-sum BP decoder is then applied to solve the reduced decoding problem $H_{otf}\cdot \textbf{e}_{otf} = \textbf{s}$. As the $H_{otf}$ matrix has a tree-like structure, the second round of BP is guaranteed to converge if a solution exists, i.e., if $\mathbf{s}$ is linearly dependent on the columns of $H_{otf}$ such that $\textbf{s} \in {\rm \textsc{image}}(H_{otf})$.

\begin{figure}
    \includegraphics[width=\linewidth]{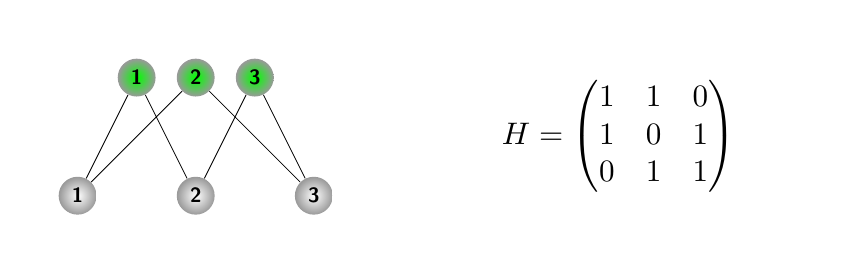}
    \includegraphics[width=\linewidth]{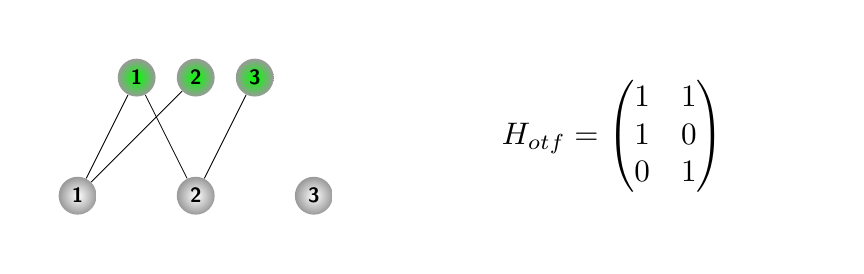}
    \caption{ Top: the Tanner graph of a classical cyclic code altogether with its parity check matrix $H$. Grey circles denote bits and green circles parity checks. Bottom: the ordered Tanner forest generated by the OTF algorithm  and its corresponding parity check matrix $H_{otf}$. Note that the third bit has been removed from the graph. This ensures that the $H_{otf}$ matrix is cycle free.    
    }
    \label{fig_loops}
\end{figure}

A loop-free graph can be obtained from any parity check matrix through the application of Kruskal's minimum spanning tree algorithm \cite{kruskal1956shortest}. This algorithm iterates through every node in the graph, eliminating those that introduce loops in the Tanner graph. An example of an OTF instance over a Tanner graph is shown in Figure \ref{fig_loops}. Note that a slight reinterpretation of the original Kruskal algorithm, as presented in \cite{kruskal1956shortest}, is necessary to adapt it to Tanner graphs. Specifically, while the original algorithm focuses on excluding certain graph edges, the modified version instead considers the data nodes of the Tanner graph. With this adjusted perspective, the modified algorithm can be implemented exactly as described in \cite{kruskal1956shortest}.

The BP+OTF decoding process can now be summarised as follows:
\begin{enumerate}
    \item Attempt to solve to the decoding using BP on the full parity check matrix. If $H\cdot\mathbf{\hat{e}}_{BP} = \textbf{s}$, return $\mathbf{\hat{e}}_{BP}$ as the solution. Else, proceed to step 2.
    \item Use the soft-information output of the BP decoding, $p_{BP}(\textbf{e})$, to order the qubits from most-to-least likely of being in the support of the error.
    \item Apply the modified Kruskal algorithm to the parity check matrix, considering qubits in the order determined in step 2, to obtain the OTF parity check matrix $H_{otf}$.
    \item Solve the OTF decoding problem using a product-sum BP decoder.
    \item Verify the output of the OTF decoding: if $H_{otf}\cdot \mathbf{\hat{e}}_{otf} = \textbf{s}$, then the decoding is valid. This will be the case for all instance of $H_{otf}$ where $\textbf{s} \in {\rm \textsc{image}}(H_{otf})$, assuming that the BP decoder has been allowed to run for a number of iterations equal to the column count of $H_{otf}$.%
\end{enumerate}

\subsection{Complexity of OTF post-processing}

The runtime complexity of each of the steps in OTF post-processing is outlined below:

\begin{enumerate}
    \item Sorting qubits by probability of error: the columns of the parity check matrix are sorted in order of decreasing probability of being in error.  Such a sort can be computed with complexity $\mathcal{O}(n\log(n))$, e.g. by means of a merge sort algorithm \cite{mergeSort}.
    \item Computing the ordered Tanner graph $H_{otf}$. The OTF can be obtained with complexity  $\mathcal{O}(n\log(n))$, using Kruskal's algorithm \cite{kruskal1956shortest} (see Appendix \ref{Appendix_UF} for more details).
    \item Belief propagation over the OTF graph. This has worst-case runtime $\mathcal{O}(n)$, where $n$ is the block-length of the code \cite{mackay1997near}. 
\end{enumerate}

The complexity is dominated by the second step with runtime $\mathcal{O}(n\log n)$. Thus, the OTF decoder therefore has almost-linear time complexity with the block-length of the code.

\subsection{Graph sparsity and OTF graph validity}\label{sec:sparsity}

The OTF post-processor can be viewed as an approximate matrix inversion method: by generating a spanning tree of the original graph, it seeks to identify a set of linearly independent columns that can reproduce the syndrome. This set of columns can be identified in $\mathcal{O}(n \log(n))$ time, which is a significant improvement over the $\mathcal{O}(n^3)$ worst-case complexity of matrix inversion via Gaussian elimination. However, unlike Gaussian elimination, Kruskal's algorithm does not guarantee that a compatible set of linearly independent columns will be found such that $\textbf{s} \in {\rm \textsc{image}}(H_{otf})$.

The success of OTF post-processing is closely tied to the sparsity of the parity-check matrix’s columns: the lower the average column weight $\bar{j}$, the more likely it is to find a valid ordered Tanner forest $H_{otf}$. To understand this, consider the process by which the OTF matrix is constructed. The search begins with an empty matrix $H_{otf}^{0}$. The OTF matrix is then populated by adding columns from the original matrix, $H$, that do not introduce loops into $H_{otf}^{i}$, where $i$ denotes the tree growth step (i.e., the number of columns added so far). At each growth step $i$, the Tanner graph of $H_{otf}^{i}$ is represented as $\mathcal{T}^{i}(V_D, V_P, E)$, where $V_D$ are the data nodes, $V_P$ are the parity nodes, and $E$ are the edges. Each column under consideration from $H$ can similarly be represented as a set of parity nodes $\Tilde{V}_P$, with the column weight $j$ defined as $|\Tilde{V}_P|$.

For a newly added column to introduce a loop in $H_{\mathrm{otf}}$, it must satisfy the condition $|\tilde{V}_P \cap V_P| > 1$. Consequently, a lower average column weight $\bar{j}$ increases the probability that a newly added column does not create a loop. Formally, assuming a regular graph with variable node degree $j$, the number of variable nodes included in an OTF graph is given by

    \begin{equation}
    n_{otf} = \frac{d-1}{j -1},
\end{equation}

where $d$ denotes the syndrome (detector) length. This expression follows from the tree condition $|V| - |E| = 1$, where $|V|$ and $|E|$ denote the number of vertices and edges, respectively. In this setting, the total number of vertices is $d + n_{\mathrm{otf}}$, while the total number of edges is $j \cdot n_{\mathrm{otf}}$, under the regularity assumption. Consequently, a sparser graph (i.e., lower column weight) yields an ordered Tanner forest $H_{\mathrm{otf}}$ with a larger number of linearly independent columns, thereby increasing the likelihood that the condition $\mathbf{s} \in \mathrm{Im}(H_{\mathrm{otf}})$ is satisfied. Furthermore, the OTF graph expansion provides a means to assess whether the number of included variable nodes exceeds the number of correctable errors. If this number is insufficient, there may exist correctable error patterns that cannot be resolved by the OTF procedure.

\section{Sparsifying detector error models}
As described in Section \ref{sec:circuit-level-decoding}, QEC codes are decoded by means of a detector graph that relates circuit fault locations to linear combinations of syndrome measurements. The detector error model enables the propagation of errors through the syndrome extraction circuit to be accounted for during decoding. Due to the richer set of error mechanisms it considers, it is typically the case that a detector error model graph will be less sparse than the corresponding CSS parity check matrices $H_{X/Z}$ of the code. This lack of sparsity is detrimental to the success of OTF post-processing, as it increases the probability that the generated OTF matrix will not satisfy the validity condition $\textbf{s} \in {\rm \textsc{image}}(H_{otf})$. See Section \ref{sec:sparsity} for more details.

We now describe a \textit{sparsification} routine which is designed to re-express a detector error model graph into a sparser form that is more suitable for OTF decoding. Specifically, our method maps the detector graph $H_{dem}\in \mathbb{F}_2^{d\times n}$ into a sparsified detector graph $H_{sdem}\in\mathbb{F}_2^{d\times n_s}$ that has maximum column weight at most equal to the maximum column-weight, $\gamma$, of $H_{X/Z}$. This is achieved by finding linear-combinations of columns of weight $\leq \gamma$ that generate each of the columns in $H_{dem}$. As the $H_{sdem}$ matrix is by design sparser and smaller than the detector matrix $H_{dem}$, applying OTF post-processor $H_{sdem}$ is more likely to result in a successful decoding.

Our mapping includes a \textit{transfer matrix} that allows the error channel associated with each error mechanism in the detector graph to be mapped to an equivalent error channel for the sparsified detector graph. This enables the soft-information output of an initial run of BP applied to the detector model to be re-purposed in the sparsified detector graph. The benefit of this is two-fold: first, the sparsified detector graph decoding will account for aspects of error propagation through the circuit, and second, the BP decoding over the OTF graph will be over a non-uniform noise channel, improving its rate of convergence.

The sparsification routine is motivated by techniques used to map detector error models for surface codes to graphs with column-weight $\leq 2$ suitable for decoding with matching decoders such as minimum-weight perfect-matching \cite{pymatching} or union-find \cite{unionfind}. Furthermore, the transfer matrix is a generalization of the methods proposed for mapping circuit-level soft-information in \cite{beliefmatching}.

We now formally define the sparsified detector error model and its associated transfer matrices.

\begin{definition}[Sparsified Detector Matrix]\label{def:Hphen}
Let the detector error model be represented by a binary matrix $H_{\mathrm{dem}} \in \mathbb{F}_2^{d \times n}$, where $H_{\mathrm{dem}} = (\mathbf{h}_{\mathrm{dem}}^1, \mathbf{h}_{\mathrm{dem}}^2, \ldots, \mathbf{h}_{\mathrm{dem}}^n)$ and each column $\mathbf{h}_{\mathrm{dem}}^i \in \mathbb{F}_2^d$. The \emph{sparsified detector matrix}, denoted $H_{\mathrm{sdem}} \in \mathbb{F}_2^{d \times n_s}$ with $n_s < n$, is defined such that 
$$
{\rm\textsc{image}}(H_{\mathrm{sdem}}) = {\rm\textsc{image}}(H_{\mathrm{dem}}),
$$
i.e., the columns of $H_{\mathrm{sdem}}$ form a generating set for the column space of $H_{\mathrm{dem}}$. Similarly, we define the \emph{sparsified logical observable matrix} $O_{\mathrm{sdem}}$, whose columns lie in the image of the original logical observable matrix $O_{\mathrm{dem}}$.
\end{definition}

Note that the selection of the sparsified detector matrix is not unique. In this work, we use the following sparsified models:
\begin{itemize}
    \item For surface codes, the standard circuit-level matching graph is used as the sparsified detector model. This is obtained by decomposing hyperedges in the full detector graph into matchable edges \cite{beliefmatching}.
    \item For bivariate bicycle codes, a \textit{quasi-phenomenological model} is used as the sparsified detector graph. This is obtained by removing the two-qubit gate errors from the circuit-level noise model.
\end{itemize}

\begin{figure*}
    \includegraphics[width = \textwidth]{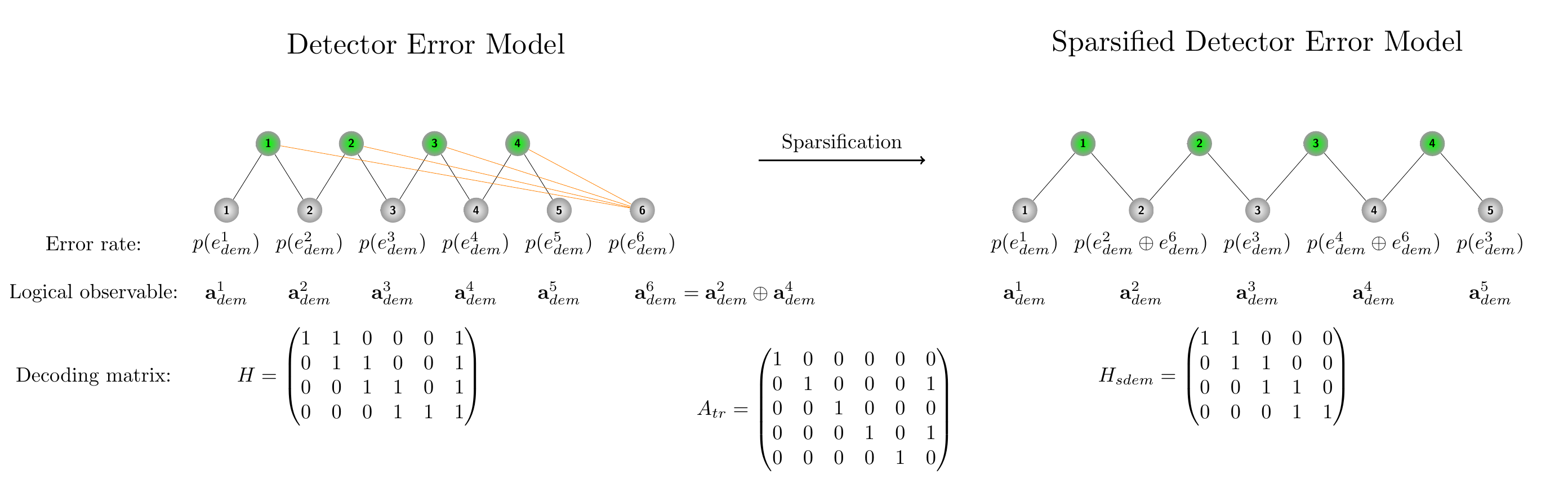}
    \caption{Sparsification process example. On the left, a detector error model, on the right, a sparsified instance of it. On the top on both sides we can see the Tanner graph of both detector error models, the orange edges denote the adjacency of the correlated variable node which will not be accounted for on the sparsified version. On the second and third rows we can see the error probabilites and the logical observables of each error in the detector error model and its sparsified version on the left and right respectively. in this particular example $\mathbf{a}_{dem}^i = O_{dem}\cdot\mathbf{e}^i_{dem}$ and $\oplus$ refers to the XOR operation. Below, the decoding parity check matrix of each error model. On the bottom, the specific example transfer matrix.}
    \label{fig:sparsified_example}
\end{figure*}

We observe that the number of columns in the sparsified detector models is typically $\approx30\%$ of the original DEM. This shows that the models are not only sparser, but also much smaller.

Recall that each column in a detector error model matrix, $H_{dem}$, corresponds to a fault mechanism in the code's syndrome extraction circuit, each of which triggers a sequence of detectors indexed by the non-zero entries in the column. As such, $H_{sdem}$ represents a reduced set of error mechanisms. We now define the transfer matrix that relates the full detector error model to its sparsified form:

\begin{definition}[The transfer matrix]
The transfer matrix $A_{tr}\in\mathbb{F}_2^{n_s\times n}$ describes the mapping from sparsified detector model to the full detector error model
\begin{equation}
H_{sdem}\cdot A_{tr} = H_{dem}.
\end{equation}
Each column $\textbf{a}^i$ of $A_{tr}$ is a vector that maps column $\textbf{h}^i_{dem}$ of $H_{dem}$ to a linear combination of the columns of $H_{sdem}$
\begin{equation}
    \mathbf{h}_{dem}^i = \sum_j a_j^i \mathbf{h}_{sdem}^j\rm.
\end{equation}
The transfer matrix preserves the action of the logical observable matrix such that
\begin{equation}
    O_{dem}\cdot \mathbf{e}_{dem}^i = O_{sdem}\cdot \left(\sum_j a_j^i \mathbf{e}_{sdem}^j\right)\rm,
\end{equation}
where $\mathbf{e}_{dem}^i$ and $\mathbf{e}_{sdem}^j$ refer to the circuit faults associated to columns $i$ and $j$ of the detector and sparsified models accordingly.

\end{definition}

Importantly, the decomposition given by the transfer matrix maps the faults of the detector error model to combinations of faults on the sparser detector graph with \textit{the same logical action}. Note, also, that the transfer matrix is not unique: multiple decompositions of the columns of $H_{dem}$ into components of $H_{sdem}$ are possible. Our primary objective is to identify sparse transfer matrices, where each column $\textbf{a}^i$ in $A_{\text{tr}}$ has the smallest possible Hamming weight. This is crucial because the transfer matrix inherently maps low-weight fault mechanisms in the detector error graph to higher-weight errors in the sparsified graph. By ensuring the transfer matrix is sparse, we minimise the likelihood that the mapped errors will exceed the code distance. Additionally, a sparse transfer matrix is advantageous for mapping soft-information to the sparsified detector graph, as will be expanded upon in the next section.

The question as how to best optimise the sparsified detector model and the transfer matrix is an interesting question for future work. For the QEC codes simulated in this work, we found that an exhaustive search method was sufficient. Details of our exhaustive strategy are outlined in Appendix \ref{app:decomposing}.

\subsection{Belief propagation decoding using the sparsified detector graph}

In this section, we propose a two stage BP+BP decoder, where BP is first applied to the full detector model graph followed by a second round of decoding on the sparsified detector graph. To this end, we outline a procedure that uses the transfer matrix to translate the soft-information output of the first round of BP into a form that can be used as the initial error channel for the second round of BP. This ensures that the decoding on the sparsified graph accounts for information about error propagation at the circuit-level. Furthermore, as the second round of BP will be supplied with a non-uniform error channel, it is more likely to converge to a solution satisfying the syndrome.

After completing the first round of BP on the full detector graph, we obtain the decoder's soft-information output, denoted as \( p_{BP}(\textbf{e}_{dem}) \). This output assigns an error probability to each fault mechanism in the detector model \( H_{dem} \). Our objective is to develop a method to translate \( p_{BP}(\textbf{e}_{dem}) \) into a corresponding probability vector \( p(\textbf{e}_{sdem}) \), which allocates error probabilities to each fault mechanism in the sparsified detector error model \( H_{sdem} \).

The transfer matrix \( A_{tr} \) establishes a relationship between fault mechanisms in the detector model and those in the sparsified detector model. Specifically, the list of detector fault mechanisms that trigger a given fault mechanism \( i \) in the sparsified model is represented by \( \{j: A_{tr}^{ij} \neq 0\} \).

When combining multiple error probabilities from the detector graph into a single probability for the sparsified graph, it is crucial to account for parity. For instance, if two individual error mechanisms in the detector error model, \( \mathbf{e}_{dem}^1 \) and \( \mathbf{e}_{dem}^2 \), trigger the same component in the sparsified detector matrix, \( \mathbf{e}_{sdem}^k \), their combined effect would be \( (\mathbf{e}_{sdem}^k + \mathbf{e}_{sdem}^k) \mod 2 = \mathbf{0} \). Therefore, probabilities should be summed only when an odd number of detector fault mechanisms contribute to the relevant sparsified detector fault mechanism. Thus, we want to determine the probability of the sparsified faults as the random variable defined as the modulo-$2$ sum (parity constraint) of the binary random variables associated to the detector error model faults that relate to it via $A_{tr}$. This mapping of probabilities from \( H_{dem} \) to \( H_{sdem} \) can then be accomplished by considering the probability of the exclusive OR (XOR) of the component binary random variables, as follows \footnote{Recall that the XOR logical operation is true if and only if the number of true inputs is odd. Furthermore, XOR and modulo-$2$ sum are equivalent for the binary field, i.e. a parity constraint. We stick to the XOR operation so as to relate the update rule to the Piling-up lemma, which is formulated by means of it \cite{piling}.}

\begin{align}\label{eq:mapLLR}\nonumber
    p(e_{sdem}^i) = 
    p\left(\bigoplus_{k\in \{j : A_{tr}^{ij}\neq 0 \}} e_{dem}^k\right) \\ = \frac{1}{2}\left(1-\prod_{k\in \{j : A_{tr}^{ij}\neq 0 \}} (1-2p(e_{dem}^k))\right),
\end{align}
where by $A_{tr}^{ij}$ we refer to element $j$ of row $i$ of the transfer matrix, $\oplus$ refers to the XOR operation and by $e_{sdem}^i$ and $e_{dem}^k$ we refer to the binary random variables associated to the sparsified and circuit faults $i$ and $k$, respectively. The combined probability in expression \ref{eq:mapLLR} can be derived by means of the Piling-up lemma that describes the probability of the XOR-clause of $n$ independent binary random variables \cite{piling}, which is the case discussed here. One can see an example of a sparsification process of a correlated repetition code in Figure \ref{fig:sparsified_example}, where a correlated variable node triggering all checks is removed when moving into the sparsified detector error model.

\begin{algorithm}
\caption{BP+BP decoder for circuit-level noise}
\label{alg:BPBP}
\begin{flushleft}
        \textbf{INPUT:} Measured syndrome $\mathbf{s}\in\mathbb{F}_2^{d}$\\
        A priori probabilities of detector error model mechanisms, $p_{ch}\in\mathbb{R}^{n}$ \\
        Detector matrix: $H_{dem}\in\mathbb{F}_2^{d\times n}$,\\
        Sparsified detector matrix: $H_{sdem}\in\mathbb{F}_2^{d\times n_s}$, \\
        Transfer matrix: $A_{tr}\in\mathbb{F}_2^{n_s\times n}$\\
        \textbf{OUTPUT:} Estimated error, $\mathbf{\hat{e}}$. 
\end{flushleft}
\begin{algorithmic}[1]
\State $(\mathbf{\hat{e}}_{dem},\  p(\mathbf{e}_{dem})) \gets$ BP\_decode($\mathbf{s}$,\ $p_{ch}$,\ $H_{dem}$)
\If{$\mathbf{s} == \mathbf{\hat{s}}_{dem}$}
    \State \textbf{return } $\mathbf{\hat{e}}_{dem}$
\EndIf
\State $p(\mathbf{e}_{sdem}) \gets$ Map($p(\mathbf{e}_{dem})$,\ $A_{tr}$)
\State $\mathbf{\hat{e}}_{sdem}\gets$ BP\_decode($\mathbf{s}$,\ $p(\mathbf{e}_{sdem})$,\ $H_{sdem}$)
\State \textbf{return } $\mathbf{\hat{e}}_{sdem}$
\end{algorithmic}
\end{algorithm}

The complete BP+BP algorithm is outlined in Algorithm \ref{alg:BPBP}. Both stages of the BP algorithm run in linear time. To fully understand the performance of the BP+BP decoder, we need to analyse the runtime scaling of the mapping from the detector error model to the sparsified model. To this end, it is insightful to treat the transfer matrix $A_{tr}$ as analogous to a Tanner graph, where the variable nodes represent elements of the detector error model, and the check nodes correspond to elements of the sparsified detector error model. Equation \ref{eq:mapLLR} can be interpreted as a message-passing update from the variable nodes to the check nodes. By using the soft information from the first BP update round as the a posteriori soft information, the probabilities $p(e^i_{sdem})$ can be computed in a single message-passing step. The computational complexity of this mapping process is essentially equivalent to that of running BP. The key difference is that the column weight is now determined by the transfer matrix, which is not uniquely defined. However, in all QEC codes examined in this study, the transfer matrices did not exhibit column weights exceeding 3, suggesting that it is generally lower than the column weight of the detector error model. Therefore, the worst-case complexity of the overall BP+BP decoder can be bounded by $\mathcal{O}(n)$, indicating linear time complexity.

\section{An almost-linear time decoder for QLDPC codes under circuit-level noise}
We now outline BP+BP+OTF as a decoding method with almost-linear runtime complexity for QLDPC codes operating under circuit-level noise. The decoder first runs the two-stage BP+BP decoder on the detector graph and its corresponding sparsified graph as described in the previous section. If BP+BP fails, the OTF decoder is applied to the sparsified graph as a post-processor. Pseudo-code for the BP+BP+OTF decoder is provided below:

\begin{algorithm}
\caption{BP+BP+OTF decoder for circuit-level noise}
\label{alg:BPBPOTF}
\begin{flushleft}
        \textbf{INPUT:} Measured syndrome: $\mathbf{s}\in\mathbb{F}_2^{d}$\\
        a priori probabilities of DEM error mechanisms, $p_{ch}\in\mathbb{R}^{n}$ \\
        Detector matrix: $H_{dem}\in\mathbb{F}_2^{d\times n}$,\\
        Sparsified detector matrix: $H_{sdem}\in\mathbb{F}_2^{d\times n_s}$, \\
        Transfer matrix: $A_{tr}\in\mathbb{F}_2^{n_s\times n}$\\
        \textbf{OUTPUT:} Estimated error, $\mathbf{\hat{e}}$. 
\end{flushleft}
\begin{algorithmic}[1]
\State $(\mathbf{\hat{e}}_{dem}, \ p(\mathbf{e}_{dem})) \gets$ BP\_decode($\mathbf{s}$,\ $p_{ch}$,\ $H_{dem}$)
\If{$\mathbf{s} == \mathbf{\hat{s}}_{dem}$}
    \State \textbf{return } $\mathbf{\hat{e}}_{dem}$
\EndIf
\State $p(\mathbf{e}_{sdem}) \gets$ Map($p(\mathbf{e}_{dem})$,\ $A_{tr}$)
\State $(\mathbf{\hat{e}}_{sdem},\  p(\mathbf{e}_{sdem})) \gets$ BP\_decode($\mathbf{s}$,\ $p(\mathbf{e}_{sdem})$,\ $H_{sdem}$)
\If{$\mathbf{s} == \mathbf{\hat{s}}_{sdem}$}
    \State \textbf{return } $\mathbf{\hat{e}}_{sdem}$
\EndIf
\State $\mathbf{\hat{e}}_{otf} \gets \text{OTF}(H_{sdem},\ p(\mathbf{e}_{sdem}))$
\State \textbf{return } $\mathbf{\hat{e}}_{otf}$
\end{algorithmic}
\end{algorithm}
The runtime complexity of this decoder is dominated by the $\mathcal{O}(n\log (n))$ complexity of the OTF post-processing stage. As such, the full BP+BP+OTF decoder is an almost-linear time decoder for circuit-level decoding of QLDPC codes.

\begin{figure*}
    \begin{center}
    \def\mathdefault#1{#1}
            \input{plots/BBcodesPerformance.pgf}
    \end{center}
    \caption{Logical error rate per syndrome extraction round as a function of the physical error rate for bivariate bicycle codes. Each code is simulated over a number of syndrome rounds equal to its distance. The shading highlights the region of estimated probabilities where the likelihood ratio is within a factor of $1,000$; similar to a confidence interval.}
    \label{fig:BBresults}
\end{figure*}

\section{Results}
We numerically benchmark the performance of the BP+BP+OTF decoder via Monte Carlo simulations of the bivariate bicycle codes and rotated surface codes under a depolarising circuit-level noise model \cite{BBcodes}. Precise details of the numerical methods used for this section can be found in Appendix \ref{app:numerical}, whilst a detailed specification of the circuit-level noise model is outlined in Appendix \ref{app:cln}.

\subsection{Bivariate Bicycle Codes}

Figure~\ref{fig:BBresults} shows the logical error rate per round for families of bivariate bicycle codes decoded using BP+OSD and BP+BP+OTF. Simulations were performed on the $[[72,12,6]]$, $[[108,8,10]]$, $[[144,12,12]]$, $[[288,12,18]]$ and $[[360,12,24]]$ bivariate bicycle codes first introduced in~\cite{BBcodes}. The decoder setup is as follows:

\begin{itemize}
    \item \textbf{BP+OSD-0:} We set the maximum number of iterations for the minimum-sum belief propagation decoder to $10{,}000$, following~\cite{BBcodes}. As noted in~\cite{AmbiguityClusteringOSD}, using a large iteration count improves speed, as premature termination means the decoder relies more on OSD-postprocessing. The `0' in OSD-0 indicates that no higher-order post-processing is applied find solutions that improve upon the initial output~\cite{Joschka_decoding_across}.
    
    \item \textbf{BP+BP+OTF:} Oscillations in the soft output of BP can emerge after a sufficient number of iterations~\cite{bpgdg}, which may result in OTF graphs that fail to span the observed syndrome. To address this, we adopt an ensemble decoding strategy~\cite{googleEnsmeble, koutsioumpas2025automorphismensembledecodingquantum}, where each member of the ensemble uses a different number of iterations in the first BP stage. This improves robustness without affecting overall runtime, as all decoders run in parallel \footnote{For the $[[72,12,6]]$ code, which exhibits limited degeneracy, we do not use the ensemble approach. Instead, we run a single BP+BP+OTF decoder with $300$ iterations on the full detector graph and $113$ iterations in the second BP stage and on the OTF graph.}. The ensemble consists of $22$ BP+BP+OTF decoders, with first-stage BP iteration counts ranging from $17$ to $391$ in steps of $17$. The [[360,12,24]] uses an ensemble of size $44$, with first-stage ranging from $7$ to $399$. The subsequent BP rounds on both the sparsified and OTF graphs are capped at $113$ iterations. All BP stages use the minimum-sum variant. See Appendix \ref{app:convergence} for a detailed discussion on why the number of iterations on the OTF stage is enough and the need for the ensembling approach.

\end{itemize}

\begin{figure*}[!ht]
\centering
\begin{subfigure}{.5\textwidth}
    \def\mathdefault#1{#1}
    \scalebox{0.57}{\input{plots/timing_p_003.pgf}}
\caption{physical error rate @ $p=0.003$.}
\end{subfigure}
\begin{subfigure}{.49\textwidth}
    \def\mathdefault#1{#1}
    \scalebox{0.57}{\input{plots/timing_p_004.pgf}}
    \caption{physical error rate @ $p=0.004$.}
\end{subfigure}%
\caption{Mean decoding time per round for BP+OSD-0 and BP+BP+OTF when decoding bivariate bicycle codes. For the BP+BP+OTF ensemble we record the time of the first instance that finishes decoding. a) Decoding at physical error rate $p=0.003$. b) Decoding at physical error rate $p=0.004$.}
\label{fig:BBresultstiming}
\end{figure*}

The results demonstrate that the proposed decoder matches the performance of BP+OSD-0 when decoding bivariate bicycle codes. Crucially, this is achieved with an almost-linear runtime, requiring at most $617$ BP iterations across the three stages of the decoder.

To evaluate the speedup provided by our implementation, we also performed timing benchmarks. Full details of these tests are provided in Appendix~\ref{app:numerical}. Figure~\ref{fig:BBresultstiming} reports the mean decoding time per syndrome extraction round for both decoders at two physical error rates, $p = 0.003$ and $p = 0.004$ (see Appendix~\ref{app:timing} for detailed statistics). The results clearly show that our decoder consistently outperforms the standard BP+OSD decoder, achieving speedups greater than a factor of $10$ across all scenarios.

This performance gap should be interpreted with care. The BP implementation used from~\cite{joschkaLDPC} executes serially, even under a parallel schedule. As a result, the actual complexity of our current implementation is quadratic. A fully parallel implementation would yield significantly greater speedups compared to BP+OSD. Nonetheless, these results already highlight that the BP+BP+OTF decoder exhibits much better scaling behaviour, particularly for codes with higher distances and larger detector error models.

Another important aspect of our analysis concerns the amount of decoding information lost when the detector error model (DEM) is sparsified. To investigate this, we conduct a series of numerical experiments using BP+OSD-0 in different configurations, with detailed results presented in Appendix \ref{app:BPBPOSD}.

First, we compare the performance of BP+OSD-0 applied directly to the full DEM with the same decoder applied to the sparsified model, both using a maximum of 10,000 BP iterations. We observe a significant degradation in performance, with logical error rates worsening by several orders of magnitude, and no improvement in decoding time. This indicates that completely ignoring the structure of the full DEM is not an effective strategy.

We then evaluate the BP+BP+OSD-0 decoder, where the first BP stage runs on the full DEM for 1,000 iterations and the second stage runs on the sparsified model for 9,000 iterations, followed by OSD-0 on the sparsified model. In this setup, 90 percent of the BP iterations and all post-processing are performed on the sparsified graph. Remarkably, this hybrid approach matches the logical error rate of BP+OSD-0 executed entirely on the full DEM, while achieving a threefold reduction in average decoding time.

These results support the effectiveness of our sparsification technique. Running a limited number of BP iterations on the full DEM is sufficient to extract the essential soft information required for decoding. Since the sparsified model is also smaller, other decoders that perform well under phenomenological noise may benefit from this approach, enabling them to operate effectively under circuit-level noise with reduced dependence on the full DEM.

\subsection{Surface codes}

\begin{figure*}[!ht]
\centering
\begin{subfigure}{.5\textwidth}
    \def\mathdefault#1{#1}
    \scalebox{0.6}{\input{plots/SCMWPMvsOTF.pgf}}
\caption{BP+BP+OTF versus MWPM.}
\end{subfigure}
\begin{subfigure}{.49\textwidth}
    \def\mathdefault#1{#1}
    \scalebox{0.6}{\input{plots/SCOSDvsOTF.pgf}}
    \caption{BP+BP+OTF versus BP+OSD-0.}
\end{subfigure}%
\caption{Logical error rate per syndrome extraction round as a function of the physical error rate for rotated surface codes. Each code is simulated over a number of syndrome rounds equal to its distance using Stim's surface\_code:rotated\_memory\_z experiment. The shading highlights the region of estimated probabilities where the likelihood ratio is within a factor of $1,000$; similar to a confidence interval.}
\label{fig:SCresults}
\end{figure*}

Figure \ref{fig:SCresults} presents the logical error rates per round for rotated surface codes with distances $d = {5, 9, 13, 17}$, decoded using minimum-weight perfect matching (MWPM), BP+OSD-0, and BP+BP+OTF. For the BP+OSD-0 decoder, we use 70 minimum-sum BP iterations, while the BP+BP+OTF decoder employs 6, 51, and 50 maximum iterations for its three decoding stages, respectively. As shown in the figure, BP+BP+OTF achieves logical error suppression comparable to MWPM (Figure \ref{fig:SCresults}a), and significantly outperforms BP+OSD-0 (Figure \ref{fig:SCresults}b). This result is particularly notable, as BP+OSD-0 is known to slightly outperform belief-find \cite{joschkaLSD}, a decoder with almost-linear runtime complexity. These findings highlight the strong performance of BP+BP+OTF in both accuracy and efficiency.

A distinctive feature of the BP+BP+OTF decoder in the context of surface codes is its guaranteed convergence. This property arises from the fact that the OTF post-processor operates on a matrix with a matching structure. In the case of surface codes without periodic boundaries, an additional virtual node must be included in the matrix to enable OTF decoding (see Appendix \ref{app:virtual}). This results in an augmented matrix with $d+1$ rows, and each column has weight 2. Since a tree-like graph must satisfy $|V| - |E| = 1$, where $|V|$ and $|E|$ are the numbers of vertices and edges, respectively, we can deduce the number of columns in the OTF matrix. Letting $n_{\text{otf}}$ denote the number of such columns, the relationship becomes $(d+1 + n_{\text{otf}}) - 2n_{\text{otf}} = 1$, which implies $n_{\text{otf}} = d$. The final BP step is then performed on a full-rank $d \times d$ matrix, excluding the virtual node, ensuring that the syndrome is always in the image of $H_{\text{otf}}$. This guarantees convergence and makes BP+BP+OTF particularly attractive for codes with matching structure \cite{decoders}.

We do not explicitly benchmark the runtime of our BP+BP+OTF implementation against \textit{pymatching} \cite{Sparse_blossom}, since \textit{pymatching} is highly optimised for CPU execution, while our BP implementation is serial and not performance-tuned. Nonetheless, it is noteworthy that BP+BP+OTF is capable of decoding a $d = 17$ surface code using a total of only 107 BP iterations. This suggests that a hardware-accelerated implementation, for example on an FPGA, could achieve decoding speeds and meet the requirements for real-time operation \cite{relayBP}.

\section{Conclusion}

In this work, we introduced BP+BP+OTF as an almost linear-time decoder for QLDPC codes under circuit-level noise. The OTF post-processor removes qubits from the decoding graph until it achieves a tree-like structure, thereby increasing the likelihood that subsequent rounds of BP will converge. Additionally, we presented a novel method for mapping detector error models to sparser matrices while preserving critical information about circuit-level fault propagation. Numerical simulations demonstrate that using this mapping, the BP+BP+OTF decoder matches the performance of state-of-the-art decoders such as BP+OSD and MWPM when applied to families of bivariate bicycle and surface codes under circuit-level noise.

The sparsification routine we propose extends beyond the BP+BP+OTF decoder. Our simulations suggest that performing an initial round of decoding on the full detector model, followed by a second round on the sparsified detector model, can enhance performance across various decoders, including plain BP and BP+OSD. The sparsified detector model features fewer loops, fewer columns, and a less redundant structure. Moreover, by mapping the soft information from the first BP round to the sparsified detector model, the second BP round is supplied with a non-uniform error channel, accelerating convergence. Our results show that this mapping to the sparsified detector model can be used to match the performance of BP+OSD-0 executed over the full DEM while significantly accelerating its speed. We anticipate that other decoder families -- for example BP+AC \cite{AmbiguityClusteringOSD}, BP+LSD \cite{joschkaLSD}, and BP+CB \cite{closedbranch} -- will also benefit from the sparsified detector error model. Furthermore, being able to operate over a detector model that resembles a phenomenological parity check matrix may make several decoding proposals to succeed at decoding circuit-level noise when those have shown promise over the simplified phenomenological noise model.

In future work, we will explore the application of the BP+BP+OTF decoder to various code families, including colour codes, hypergraph product codes \cite{xu2023constantoverheadfaulttolerantquantumcomputation}, and lifted product codes \cite{scruby2024highthresholdlowoverheadsingleshotdecodable}. 

The primary failure mode of the BP+BP+OTF algorithm arises in cases where the OTF post-processor fails to produce a spanning tree capable of supporting the syndrome, i.e., when $\textbf{s} \notin {\rm \textsc{image}}(H_{otf})$. The probability of such failures can be reduced by optimising the sparsity of the transfer matrix that maps the detector model to its sparsified form. In this work, we used an exhaustive approach to derive the mapping, but further optimisations could be achieved by, for example, explicitly excluding elements of $H_{sdem}$ that introduce loops. Furthermore, the pre-processing step for computing the transfer matrix is carried out via an exhaustive search over a local radius (see Appendix \ref{app:decomposing}). While this approach is efficient for the qLDPC codes considered in this work, it may become intractable for future systems involving tens of thousands of qubits. More scalable alternatives could be developed using Pauli propagation frameworks such as Stim, or by solving the corresponding linear systems via matrix inversion techniques. These remain an area for future investigation.

Concurrently with our work, a number of high-performance decoders have been proposed, including the beam search decoder~\cite{beam_search}, the GARI decoder~\cite{gari}, the Vibe Decoder \cite{koutsioumpas2025colourcodesreachsurface}, and the Relay-BP decoder~\cite{relayBP}. In particular, Relay-BP is a message-passing algorithm that is well suited to FPGA implementation and exhibits linear-time complexity. However, the number of iterations required for convergence is not bounded a priori, leading to latency tails in which certain syndromes may require a large number of decoding steps~\cite{relay_fpga}. Moreover, its performance depends on careful tuning of memory parameters, which can be challenging and may vary across noise conditions~\cite{relayBP}. In contrast, the OTF procedure provides a bounded decoding stage whenever $\mathbf{s} \in \mathrm{Im}(H_{\mathrm{otf}})$, and does not rely on parameter optimisation. Furthermore, it uses standard BP subroutines, opening the possibility of using existing, and highly optimised, \textit{off-the-shelf} FPGA implementations. At the same time, we anticipate that improved heuristics from approaches such as Relay-BP or GARI could be leveraged to more effectively identify OTF graphs that explain the observed syndrome. In this sense, these recent developments are naturally complementary to our proposal.

Following this line of investigation, we observed that oscillations in the a posteriori information produced by BP can be detrimental, as they may prevent the OTF step from yielding a solution consistent with the observed syndrome. To address this issue, we adopted an ensemble-based approach that enabled reliable decoding of bivariate bicycle codes. While this process significantly improves the performance of the decoder, it is worth noting that it leverages parallel computational resources to generate multiple OTF graph candidates for a given syndrome. On modern hardware platforms such as GPUs and FPGAs, this parallelism can be exploited efficiently, often reducing the typical time-to-solution through early termination once a valid solution is found. As a result, the overall energy consumption per decoded syndrome can remain comparable to that of a single decoder requiring more sequential iterations. Future work could explore alternative strategies to improve the stability and reliability of the soft information. For example, automorphism ensemble BP \cite{koutsioumpas2025automorphismensembledecodingquantum}, modified versions of BP \cite{kuoDegen,relayBP}, or techniques such as iteration buffers \cite{bpgdg,DTD} could increase the likelihood of convergence. 

In addition, careful optimisation of the number of BP iterations performed on both the full detector error model and the sparsified model is essential for achieving the best performance with BP+BP+OTF. In this work, we explored iteration schedules that match the decoding performance of state-of-the-art methods. Further improvements may be possible through dynamic halting strategies, where the number of iterations is adapted based on the evolution of the a posteriori information or the density of the observed syndrome.

Further improvements in the runtime of BP+BP+OTF could be achieved by exploring parallelisation methods for the tree search step in the OTF post-processor. One possible approach is to combine OTF post-processing with the parallel cluster-growth strategies employed by the BP+LSD decoder \cite{joschkaLSD}.

The OTF post-processor operates on the principle that QLDPC decoding can be enhanced by modifying the structure of the decoding graph. Specifically, Kruskal's algorithm is used to identify and eliminate variable nodes that introduce problematic cycles in the Tanner graph. Similarly, the recently introduced BP+GD decoder iteratively modifies the decoding graph by excluding variable nodes with the least uncertainty in their soft-information \cite{yao2024beliefpropagationdecodingquantum}. An interesting direction for future research would be to explore the combination of BP+OTF and BP+GD.

Given its low complexity, BP+BP+OTF is a promising candidate for real-time decoding of syndromes from experimental quantum computers. To this end, dedicated hardware implementations of the algorithm using FPGAs or ASICs will be necessary. Since the BP+BP+OTF algorithm uses standard methods like the well-established minimum-sum implementation of BP and Kruskal's minimum spanning tree algorithm, it may be possible to construct such a decoder by combining existing commercially available chips. Consequently, development costs could potentially be lower compared to more specialised decoders such as BP+OSD \cite{9562513}.\\

\section{Code availability}
The code for the BP+OTF decoder can be found in the following Github repository: \url{https://github.com/Ima96/BPOTF}. In future versions we will include the implementation of the ensemble BP+BP+OTF decoder.

\section{Acknowledgements}

We thank Oscar Higgott for many useful comments and recommendations, as well as Anqi Gong for discussions on loops and trapping sets present in the detector error models of bivariate bicycle codes. We also thank Pedro Crespo for his guidance and the other members of the Quantum Information Group at Tecnun for their support.

\section{Funding}

This work was supported by the Spanish Ministry of Economy and Competitiveness through the MADDIE project (Grant No. PID2022-137099NBC44), by the Spanish Ministry of Science and Innovation through the project Few-qubit quantum hardware, algorithms and codes, on photonic and solid state systems (PLEC2021-008251), and by the Ministry of Economic Affairs and Digital Transformation of the Spanish Government through the QUANTUM ENIA project call - QUANTUM SPAIN project, and by the European Union through the Recovery, Transformation and Resilience Plan - NextGenerationEU within the framework of the Digital Spain 2026 Agenda. J.E.M. also acknowledges support from the Basque Quantum (BasQ) strategy.

J.R. is funded by an EPSRC Quantum Career Acceleration Fellowship, grant code UKRI1224. J.R. also acknowledges the support of EPSRC grants EP/T001062/1 and~EP/X026167/1.

\section{Author Contributions}
A.deM.iO. and J.E.M. conceived the research. A.deM.iO., J.R. and J.E.M. proposed the OTF algorithm. A.deM.iO., I.E.M. and J.E.M. programmed the software implementing the decoding algorithm. A.deM.iO. and J.E.M. conceived the sparsification technique and derived the transfer matrix rules. A.deM.iO., I.E.M. and J.E.M. conducted the numerical simulations. All the authors contributed in analysis the data and writing the manuscript. J.E.M. and J.R. obtained their corresponding funding sources.

\section{Competing Interests}
The authors declare no competing interests.

\bibliographystyle{IEEEtran}
\bibliography{bibliography}

\begin{appendix}

\section{The Union-Find data structure for the OTF decoder} \label{Appendix_UF}

In order to search an OTF in almost-linear time, we need to consider the union find data structure. Let us consider a dynamic list of a length equal to the number checks in the code to decode, where each element is a 2-element list, we shall name that list as DYNAMIC\_LIST. The first element of each duple will indicate the root of the element and the second one will indicate its depth. At the beginning, each element $i$ will be its own root and its depth will be $1$, i.e. DYNAMIC\_LIST$\left[i\right] = [i,1]$. Each element in the dynamic list will become a tree of a single element. Afterwards, we will consider columns in the parity check matrix in the order established by the posterior probabilities. Once we consider a column in the parity check matrix, we will study its non-trivial elements. Considering $j$ non-trivial elements, we will search at the root of each of them in the DYNAMIC\_LIST.  If the $j$ roots are different, introducing the edges emanating from the variable node which represents the column will not produce a loop in the Tanner graph. If that is the case, all the elements will be merged into a single tree and the root with the largest depth will become the root of the overall tree. Subsequently, the remaining roots will change their first element in the DYNAMIC\_LIST to be the new root. Moreover, if there are two or more roots with largest depth, one will be chosen arbitrarily to be the new root of the merged tree and its depth will be increased by one.

\begin{figure}
    \centering
    \includegraphics[width=0.75\linewidth]{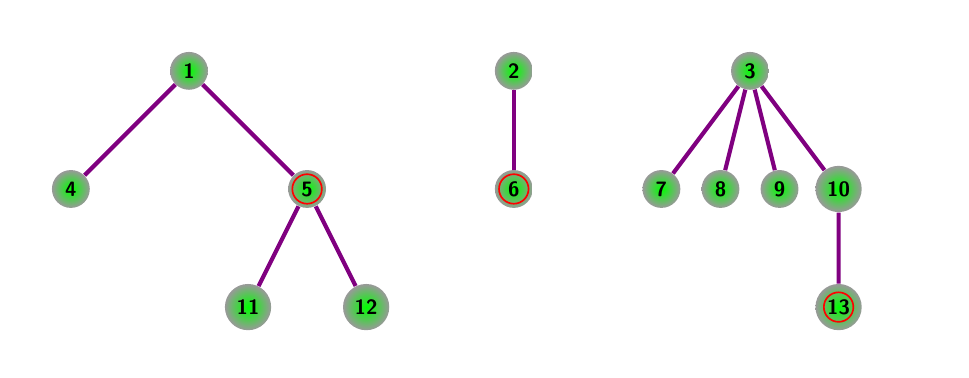}
    \includegraphics[width=0.75\linewidth]{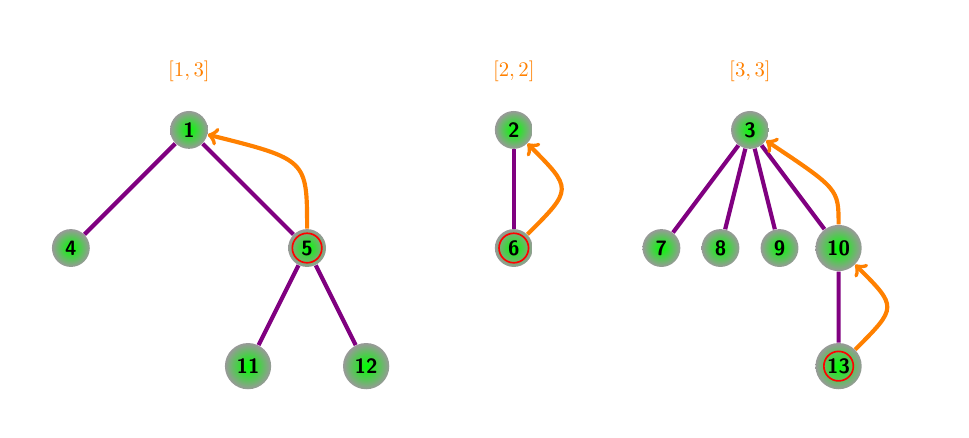}
    \includegraphics[width=0.75\linewidth]{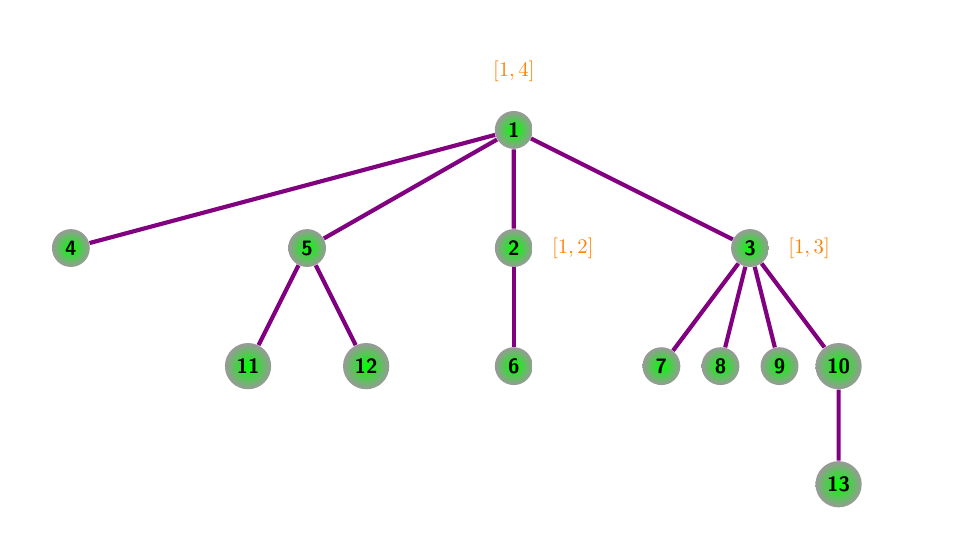}
    \caption{Process of considering a column in the parity check matrix for the ordered Tanner forest. Green circles represent the elements of the DYNAMIC\_LIST, which are labelled. For a violet line connecting a node on the bottom $j$ with a bottom on the top $i$, DYNAMIC\_LIST$\left[j\right]\left[0\right] = i$. The \textbf{top} figure indicates three trees with three different roots on the top. The three red-circled nodes belong to the non-trivial elements of the column that is being considered. On the \textbf{middle} image, the value of the three roots of each considered node is found within the DYNAMIC\_LIST. Since all roots are different, we can incorporate the column, as it does not produce a loop. We choose the root of the tree on the left as the overall root and, in the \textbf{bottom} figure, we merge the three trees by changing the first element in the DYNAMIC\_LIST of the elements 2 and 3 to be 1. Moreover, as the tree on the left and the one on the right had the same depth, the depth of the overall tree increases by 1 to be 4.}
    \label{UF}
\end{figure}

After the first step, there will be elements in the DYNAMIC\_LIST for which the first element will not satisfy DYNAMIC\_LIST$\left[i\right]\left[0\right] = i$, but the new root that it has adopted once merging with other trees. Therefore, if we want to know its root, we will have to search the first element of the dynamic list of each element that we find until the condition DYNAMIC\_LIST$\left[i\right]\left[0\right] = i$ is satisfied. Figure \ref{UF} portrays an example of the process of considering a column with three non-trivial elements, which results in the merging of three trees.

The objective of the OTF decoder will be to consider all columns from the parity check matrix following this rationale. After all the columns have been considered, the ones that will have been kept for the OTF matrix will not produce loops in the Tanner graph. Ultimately, this process will involve considering $n$ columns from the parity check matrix, for each column to consider $j$ non-trivial elements, and for each of these elements to look for their root, which can be done in, at most, $\log(n)$ time. Therefore, the overall worst-case complexity of the process of searching an OTF given a parity check matrix is $\mathcal{O}(nj\log(n))$. Due to the fact that the weight of the column is assumed to be constant when increasing block length, the overall complexity can be expressed as $\mathcal{O}(n\log(n))$.

\section{Finding the transfer matrix by means of an exhaustive search}\label{app:decomposing}
\begin{algorithm}[h!]
\caption{Decomposing detector error model detectors into ``phenomenological'' elements}
\label{alg:decomposition}
\begin{flushleft}
        \textbf{INPUT:} detector error model column $\mathbf{h}_{dem}^i\in\mathbb{F}_2^d$,\\
        ``phenomenological'' parity check matrix $H_{sdem}=(\mathbf{h}_{sdem}^1,\cdots,\mathbf{h}_{sdem}^{n_s})\in\mathbb{F}_2^{d\times n_s}$, \\
        logical observable detector error model matrix $O_{dem}\in\mathbb{F}_2^{k\times n}$, \\
        logical observable ``phenomenological'' matrix $O_{sdem}\in\mathbb{F}_2^{k\times n_s}$\\
        \textbf{OUTPUT:} decomposition vector for the detector error model column, $\mathbf{a}^i\in\mathbb{F}_2^{n_s}$
\end{flushleft}
\begin{algorithmic}[1]
\State \Comment{Get columns of $H_{sdem}$ that at least share a non-trivial element with the detector error model column, $\mathbf{h}_{dem}^i$. The $map$ variable contains which elements of $H_{sdem}$ are retained so that the result for $H_{reduced}$ can be mapped back.}
    \State $H_{reduced} \gets \emptyset$
    \State $O_{reduced} \gets \emptyset$
    \State $map \gets \emptyset$
    \For{$j \gets 1$ \textbf{to} $n_s$}
        \For{$\text{k} \gets 1$ \textbf{to} $d$}
            \If{$\mathbf{h}_{dem}^i(k) == \mathbf{h}_{sdem}^j(k) \And \mathbf{h}_{dem}^i(k) == 1$}
                    \State $H_{reduced} \gets H_{reduced} \cup \mathbf{h}_{sdem}^j$
                    \State $O_{reduced} \gets O_{reduced} \cup \mathbf{o}_{sdem}^j$
                    \State $map \gets map \cup \{j\}$
                    \State \algorithmicbreak
            \EndIf
        \EndFor
    \EndFor
\State \Comment{Exhaustive search over the reduced matrix $H_{reduced}$ to decompose the detector error model column, $\mathbf{h}_{dem}^i$. $cols_r$ refers to the number of columns of $H_{reduced} = (\mathbf{h}_{reduced}^1,\cdots,\mathbf{h}_{reduced}^{cols_{r}})$. Vectors $\mathbf{e}_{dem}^i,\mathbf{e}_{reduced}^m$ refer to the error vectors associated to the $i$ and $m$ columns of their respective parity check matrix.}
\State $\mathbf{a}^i \gets \emptyset$
\For{$j \gets 1$ \textbf{to} $cols_{r} $}
    \State $combs \gets $ GetCombinations($[1,2,\cdots,cols_{r}], j$)
    \For{$k \gets 1$ \textbf{to} $\binom{cols_r}{j}$}
        \State $\mathbf{dec} \gets \sum_{m\in combs(k)} \mathbf{h}_{reduced}^m $
        \State $\mathbf{e}_{dec} \gets \sum_{m\in combs(k)} \mathbf{e}_{reduced}^m $
        \State $\mathbf{l}_{sdem} \gets O_{reduced}\cdot \mathbf{e}_{dec} $
        \State $\mathbf{l}_{dem} \gets O_{dem}\cdot \mathbf{e}_{dem}^i $ 
        \If{($\mathbf{dec} == \mathbf{h}_{dem}^i \And \mathbf{l}_{sdem} == \mathbf{l}_{dem}$}
            \State $\mathbf{a}^i = ones(map(combs(k)))$
            \State \algorithmicbreak
        \EndIf
        \If{$\mathbf{a}^i \neq \emptyset$}
            \State \algorithmicbreak
        \EndIf
    \EndFor
\EndFor
\State \textbf{return} $\mathbf{a}^i$
\end{algorithmic}
\end{algorithm}

We aim the decomposition of each column of the detector error model into a minimum amount of sparsified components by means of the exhaustive search shown in Algorithm \ref{alg:decomposition}. The algorithm starts by identifying the columns of $H_{sdem}$ that at least share a non-trivial component with the selected detector error model column, saved in the $H_{reduced}$ matrix. This is done to reduce the amount of elements to be considered in the exhaustive search. Once this is done, the algorithm starts to test if the detector vector in question can be expressed as a combination of the columns of the sparsified parity check matrix, starting from a single element in the sum and progressively increasing the amount of terms in the decomposition. This is done in the for loop of starting at line $17$. The GetCombinations($[1,2,\cdots,cols_{r}], j$) method creates a matrix containing all possible combinations with size $j$ of the elements of the vector $[1,2,\cdots,cols_{r}]$ as its rows. When we write $combs(k)$, we refer to getting a row of such matrix, i.e. a possible combination of columns with indexes given by such row. Therefore, the matrix obtained has size $ \binom{cols_r}{j} \times j$, and we use those combinations to test the decomposition with $j$ elements and increase such value if no decomposition is found. We also impose that a decomposition candidate must coincide in its logical effect. Once such a decomposition is found, we generate the decomposition vector $\mathbf{a}^i$ with the method $ones(map(combs(k)))$ which adds ones at locations $map(combs(k))$, i.e. at the positions found but mapped back from the $H_{reduced}$ to the $H_{sdem}$ of interest. Note that since we exhaustively search from a minimum amount of columns to a bigger one, the obtained decomposition is expected to be consisted of a minimum amount of sparsified mechanisms.

Algorithm \ref{alg:decomposition} is then looped for all the columns of the detector error model in question and a transfer matrix $A_{tr}$ is found, relating the detector error model and the sparsified parity check matrix of interest. The exhaustive search approach seems to be demanding due to the fact that the combinatorial number $\binom{cols_r}{j}$ increases very fast with $j$. Importantly, this is done in a pre-processing stage and has no impact in the latency of the decoder. Note also that detector error model matrices have a periodic structure from round to round, as it can be seen in Figure 1 of \cite{bpgdg}, implying that the processing can be done for a single propagation round and then extrapolate the rest of the transfer matrix \footnote{Being strictly correct, the first round is special due to the fact that there is no propagation of previous rounds and, thus, such first round should be processed independently \cite{bpgdg}.}. It is not within the scope of this work to optimize this decomposition, but to show that doing this kind of reduction to a sparsified noise model is beneficial to decode over circuit-level noise. Aiming at the best decomposition and finding better ways to search for those, e.g. using stabiliser simulators such as Stim, is considered future work.

\section{Convergence of belief propagation in the OTF stage for bivariate bicycle codes}\label{app:convergence}
As discussed in Section \ref{sec:sparsity}, a necessary condition for successful decoding is that the generated OTF graphs expand sufficiently to include a number of variable nodes exceeding the number of correctable errors for the bivariate bicycle codes. Table \ref{tab:bbexpansion} reports the OTF expansion alongside the code distance for the BB codes considered in the code-capacity setting. For these codes, the syndrome length typically equals the code distance, and the Tanner graphs are regular with column weight $j=3$.

\begin{table}[h!]
\centering
\caption{BB codes considered and the OTF expansions for the code capacity setting.}
\label{tab:bbexpansion}
\begin{tabular}{|c|c|}
\hline
\textbf{BB code} & $n_{otf}$   \\ \hline
$[[72,6,6]]$ & $35$   \\
$[[108,8,10]]$ & $53$   \\
 $[[144,12,12]]$ & $71$    \\
 $[[288,12,18]]$ & $143$   \\ 
  $[[360,12,24]]$ & $179$   \\ 
\hline
\end{tabular}
\end{table}

From Table \ref{tab:bbexpansion}, it is evident that the OTF trees include significantly more variable nodes than the number of correctable errors. Since this family of codes does not exhibit linear distance scaling, the gap between the number of included nodes and the code distance increases with the code length. This observation provides a simple criterion to assess when OTF graphs can include a sufficient number of variable nodes. The extension to sparsified detector error models for circuit-level noise is straightforward: the check weights increase with the code distance (and are typically larger than in the code-capacity setting, depending on the syndrome extraction circuit), while the column weights remain constant under the phenomenological model. However, in Table \ref{tab:demExp} we report such a case. Due to the fact that these graphs are irregular, we report the worst-case (highest weight) and average expansion values. Furthermore, we do so for the full DEMs and the sparsified DEM.

\begin{table*}[t!]
\centering
\caption{BB codes considered and the OTF expansions for the circuit-level noise setting. Due to graph irregularity we show worst-case and mean expansions.}
\label{tab:demExp}
\begin{tabular}{|c|c|c|c|c|}
\hline
\textbf{BB code} & Worst $n_{otf}$ DEM & Worst $n_{otf}$ SDEM & Mean $n_{otf}$ DEM  & Mean $n_{otf}$ SDEM\\ \hline
$[[72,6,6]]$ & $50$  & $125$ & $120$ & $163$ \\
$[[108,8,10]]$ & $118$  & $296$ & $283$ & $389$ \\
 $[[144,12,12]]$ & $187$   & $467$ & $447$ & $615$ \\
 $[[288,12,18]]$ & $547$  & $1367$ & $1308$ & $1807$ \\ 
  $[[360,12,24]]$ & $899$  & $2249$ & $2152$ & $2979$ \\ 
\hline
\end{tabular}
\end{table*}

From the table, we can observe that the OTFs will have enough variable nodes for the circuit noise models as well. Furthermore, it can also be seen that the expansions in the sparsified graph improve significantly relative to the full DEM, indicating that it is more convenient for OTF to work over such sparsified model. We conjecture that for all practical codes, i.e. small check weights, the number of variable nodes will always exceed the number of correctable errors. The proof of this conjecture beyond the scope of this paper and is left as future work.

\begin{figure*}[!ht]
\centering
\begin{subfigure}{.5\textwidth}
    \def\mathdefault#1{#1}
    \scalebox{0.5}{\input{plots/conv144.pgf}}
\caption{Gross code.}
\end{subfigure}
\begin{subfigure}{.49\textwidth}
    \def\mathdefault#1{#1}
    \scalebox{0.5}{\input{plots/conv288.pgf}}
    \caption{double Gross code.}
\end{subfigure}%
\caption{Convergence percentages for the Gross code and the double Gross code. The light bars show the percentage of OTF convergence considering all OTF calls. The dark bars show convergence of OTF within the decoding ensembles decoding.}
\label{fig:conv}
\end{figure*}

Having established that the OTF graphs can include sufficiently many variable nodes, we now consider whether the final BP stage (applied over the OTF graph) may fail to converge even when the constructed OTF contains a valid solution to the observed syndrome. For cycle-free graphs, the number of BP iterations required for convergence is bounded by half the diameter of the factor graph, i.e. $\lceil r/2 \rceil$~\cite{BPconvergence}, where $r$ denotes the length of the longest path. The OTF graphs are constructed via a union-find data structure using weighted quick-union, in which shallower trees are attached to deeper ones. This guarantees that the resulting trees have height at most logarithmic in the number of nodes~\cite{algorithmsUF}. Since the diameter satisfies $r \leq 2\Delta$, where $\Delta$ is the tree height, the number of required iterations can be bounded as
\[
I \leq \left\lceil \frac{r}{2} \right\rceil \leq \Delta \leq \log(d + n_{\mathrm{otf}}) \leq \log(2d).
\]
This shows that a number of BP iterations logarithmic in the graph size is sufficient to ensure convergence whenever the OTF graph explains the observed syndrome. In practice, this bound is conservative, and the number of iterations could be reduced further to improve performance. This behaviour is also supported empirically in the surface code setting considered, where no instances of non-convergence were observed (and where an OTF explaining the syndrome is guaranteed to exist).

Finally, we analyse the ``syndrome coverage'' of the generated forests, defined as the proportion of OTF graphs that successfully explain the observed syndromes. This coverage depends strongly on the heuristic used to order the columns from which OTF graphs are constructed. In our implementation, we use the minimum-sum variant of BP with a parallel schedule as the ranking heuristic. It is well known that the corresponding log-likelihood ratios exhibit significant oscillations across iterations~\cite{gitGuided}. Rather than being detrimental, this behaviour naturally produces a diverse set of soft-information outputs, which we exploit to construct an ensemble of candidate orderings. This diversity is key to generating multiple distinct OTF graphs, thereby substantially increasing the likelihood that at least one instance successfully explains the observed syndrome.

\begin{table*}[t!]
\centering
\caption{Convergence rates of OTF for the Gross code.}
\label{tab:conv144}
\begin{tabular}{|c|c|c|c|c|c|}
\hline
$\mathbf{p}$&\textbf{\# convergence} & \textbf{\# total} & \textbf{Total \%} & \textbf{\# within ensemble} & \textbf{Within \%}    \\\hline
$0.003$&$32287$ & $69890$ & $46.2\%$ & $33000$ & $97.85\%$  \\\hline
$0.002$&$9127$ & $12901$ & $70.75\%$  &  $9204$   & $99.16\%$  \\\hline
$0.001$&$1987$ & $2121$ & $93.7\%$  & $1989$ & $99.9\%$ \\
\hline
\end{tabular}
\end{table*}

\begin{table*}[t!]
\centering
\caption{Convergence rates of OTF for the double Gross code.}
\label{tab:conv288}
\begin{tabular}{|c|c|c|c|c|c|}
\hline
$\mathbf{p}$&\textbf{\# convergence} & \textbf{\# total} & \textbf{Total \%} & \textbf{\# within ensemble} & \textbf{Within \%}    \\\hline
$0.003$&$94130$ & $124673$ & $75.5\%$ & $94572$ & $99.5\%$  \\\hline
$0.002$&$28300$ & $30342$ & $93.3\%$  &  $28318$   & $99.94\%$  \\\hline
$0.001$&$7545$ & $7677$ & $98.3\%$  & $7545$ & $100\%$ \\
\hline
\end{tabular}
\end{table*}

To quantify this effect, we perform a heuristic study of convergence over generated OTF graphs for the Gross ($[[144,12,12]]$) and double Gross ($[[288,12,18]]$) codes. We simulate $10^6$ shots in the waterfall regime ($p=0.001, 0.002, 0.003$) and define two convergence ratios. The first measures the fraction of OTF graphs for which BP+BP+OTF converges at the OTF stage, relative to the total number of generated OTF graphs. The second measures the ratio between the number of OTF convergence events and the total number of decoding attempts using OTF, where each decoder instance in the ensemble is counted separately. The latter metric demonstrates that, when ensemble decoding is employed, non-convergent OTF graphs occur only rarely.

The results are shown in Figure \ref{fig:conv}. While a significant fraction of individual OTF attempts fail to converge, this is consistent with the fact that many generated OTF graphs do not explain the observed syndrome. However, when considering the full ensemble, failures of the OTF stage are extremely rare, indicating that the overall syndrome coverage is high. These results also show that convergence improves at lower physical error rates and for larger code sizes. For clarity, we summarise the numerical results in Tables \ref{tab:conv144} and \ref{tab:conv288}.

Overall, these findings demonstrate that the ensembling strategy is essential to compensate for the variability of the BP-based heuristic, ensuring that non-convergence events remain rare in practice.

\section{Numerical simulations}\label{app:numerical}

Monte Carlo computer simulations of the bivariate bicycle codes have been performed with the objective of obtaining the performance curves (logical error rate). The circuit-level noise simulations have been done the following way. The sampling of the errors arising due to the noisy stabiliser circuit noise has been done by means of Stim \cite{stim}. Stim considers the check measurements upon a set of syndrome extractions altogether with a final measurement of the data qubits. We consider $d_c$ syndrome extraction rounds, where $d_c$ is the distance of the QEC. We also use the SlidingWindowDecoding \cite{bpgdg} Github repository for constructing the Stim syndrome extraction circuits and obtain the circuit-level noise parity check and observable matrices of the bivariate bicycle codes. The decoder uses those to resolve the syndrome and return an error, which is later compared to the logical action of the Stim error.

The operational figure of merit we use to evaluate the performance of these quantum error correction schemes is the Logical Error Rate per syndrome cycle, i.e. the probability that a logical error has occurred after the recovery operation per syndrome extraction round \cite{BBcodes}.

Regarding the software implementations of the decoders used perform the numerical simulations have been: the BP+OSD implementation in the LDPC Github package \cite{Joschka_decoding_across,joschkaLDPC} (with slight modifications for handling circuit-level noise \cite{BBcodes}) and our implementation for the proposed decoder can be found in the following Github \url{https://github.com/Ima96/BPOTF}.

\begin{figure*}[!ht]
\centering
\begin{subfigure}{.5\textwidth}
\def\mathdefault#1{#1}
    \scalebox{0.57}{\input{plots/OTFstatistics_p_003.pgf}}
\caption{BP+BP+OTF at $p=0.003$.}
\end{subfigure}%
\begin{subfigure}{.49\textwidth}
    \def\mathdefault#1{#1}
    \scalebox{0.57}{\input{plots/OSDstatistics_p_003.pgf}}
    \caption{BP+OSD-0 at $p=0.003$.}
\end{subfigure}
\begin{subfigure}{.5\textwidth}
\def\mathdefault#1{#1}
    \scalebox{0.57}{\input{plots/OTFstatistics_p_004.pgf}}
\caption{BP+BP+OTF at $p=0.004$.}
\end{subfigure}%
\begin{subfigure}{.49\textwidth}
    \def\mathdefault#1{#1}
    \scalebox{0.57}{\input{plots/OSDstatistics_p_004.pgf}}
    \caption{BP+OSD-0 at $p=0.004$.}
\end{subfigure}%
\caption{Runtime distributions of BP+BP+OTF and BP+OSD-0 for the $[[288,12,18]]$ bivariate bicycle code. a) and c) show BP+BP+OTF at $p=0.003$ and $p=0.004$, respectively. b) and c) show BP+OSD-0 at $p=0.003$ and $p=0.004$. The data is separated into time-bins with width that grows logarithmically with decoding time.}
\label{fig:BBstatistics}
\end{figure*}

We also use a decimation parameter to determine how aggressively the OTF is decimated from the sparsified detector model. For bivariate bicycle codes, we observed that leaving a small probability for the nodes that do not belong to the OTF improves convergence and, thus, we set said decimation parameter at $1e-9$. For surface codes, convergence is assured over the OTF graph, which implies that the best scenario is to set the parameter to $0$ and fully decimate the sparsified detector model.

Furthermore, we observe that it is important to clip the propagated values of posteriori information from the DEM to the sparsified detector model for the sake of numerical stability. In this work, we set the clipping range at $[10^{-80},1.0-10^{-80}]$.

Regarding the timing analysis, we decode $20,000$ shots using both BP+OSD-0 and BP+BP+OTF (the same shots for both of them) and measure the time required to obtain a recovery. For the ensemble version of BP+BP+OTF we define the decoding time as the time required by the first decoder of the ensemble that reaches a solution because the ensemble can be run in parallel. All simulations have been run single-threaded on an Intel Core i9 based laptop.

\section{Circuit-level noise model}\label{app:cln}
We consider the standard depolarizing (unbiased) circuit-level noise model \cite{decoders,beliefmatching,BBcodes,bpgdg} that consists of:
\begin{itemize}
    \item \textbf{Noisy single qubit gates:} those are followed by a Pauli operator, $\{\mathrm{I,X,Y,Z}\}$, sampled independently with probabilities $p_\mathrm{X}=p_\mathrm{Y}=p_\mathrm{Z} = p/3$ and $p_\mathrm{I} = 1-p$.
    \item \textbf{Noisy two-qubit gates:} those are followed by a two-qubit Pauli operator, $\{\mathrm{I,X,Y,Z}\}^{\otimes 2}$, sampled independently with probability $p/15$ for the non-trivial operators and $p_\mathrm{I^{\otimes 2}} = 1-p$.
    \item \textbf{State preparation:} state preparations are followed by a Pauli operator which may flipped the state to its orthogonal one with probability $p$. Note that this reduces to substituting the preparation of the $\ket{0}$ state by $\ket{1}$ and the preparation of the $\ket{+}$ state by $\ket{-}$, each with probability $p$.
    \item \textbf{Measurements:} the outcomes of measurements are flipped with probability $p$.
    \item \textbf{Idle locations:} those are followed by a Pauli operator, $\{\mathrm{I,X,Y,Z}\}$, sampled independently with probabilities $p_\mathrm{X}=p_\mathrm{Y}=p_\mathrm{Z} = p/3$ and $p_\mathrm{I} = 1-p$.
\end{itemize}

\section{Statistics of the timing results}\label{app:timing}
In the main text, we presented the mean average time of BP+BP+OTF and BP-OSD-0 to show the speed improvement that our implementation of the decoder offers. However, the statistics of the decoding times are also interesting for elucidating the behavior of both decoders. Figure \ref{fig:BBstatistics} shows the histograms of the decoding times obtained from $20,000$ shots for the $[[288,12,18]]$ bivariate bicycle code. We do so for the physical error rates $p=0.003$ and $p=0.004$. We select the bigger bivariate bicycle code due to the fact that it is the most demanding code.

In such figure, we can observe that BP+OSD-0 shows a clear bimodal distribution (note the peak at the right of Figures \ref{fig:BBstatistics}b and \ref{fig:BBstatistics}d) as discussed in \cite{DTD}. The distributions for the BP+BP+OTF show to be more complex, with a couple of peaks. The reason for this is the fact that we used an ensemble version of the decoder to mitigate the negative effects of the oscillations of BP a posterioris. However, the spread in this case is much smaller than that shown by BP+OSD-0, indicating that the decoder is faster not only in average, but in all cases. Finally, an important thing here is that the BP implementation we used runs serially, as commented on in the main text. Thus, a fully parallelized implementation would make the distributions for BP+BP+OTF to drift to the left as the full decoder is based on BP. On the contrary, for BP+OSD-0 such an implementation would make the timing associated to BP to drift left leaving the OSD-0 peak unchanged.

\begin{figure*}[!ht]
\centering
\begin{subfigure}{.5\textwidth}
\def\mathdefault#1{#1}
\scalebox{0.6}{\input{plots/BBcodesPerformanceBPOSDsparse.pgf}}
\caption{BP+OSD-0 over DEM vs BP+OSD-0 over sparse model.}
\end{subfigure}
\begin{subfigure}{.49\textwidth}
    \def\mathdefault#1{#1}
    \scalebox{0.6}{\input{plots/BBcodesPerformanceBPBPOSD.pgf}}
    \caption{BP+OSD-0 over DEM vs BP+BP+OSD-0.}
\end{subfigure}%
\caption{Performance of BP+OSD-0 when operating over the sparsified detector model. a) We compare the standard BP+OSD-0 running over the full DEM with BP+OSD-0 only over the sparsified detector model. Several orders of magnitude are lost in logical error. b) We compare the standard BP+OSD-0 running over the full detector error model with the two-stage BP and an OSD-0 post-processor, i.e. BP+BP+OSD-0. $90\%$ of BP iterations are run on the sparsified model. We observe matching accuracy. The shading highlights the region of estimated probabilities where the likelihood ratio is within a factor of $1,000$; similar to a confidence interval.}
\label{fig:BPOSDsparse}
\end{figure*}
\section{Running BP+OSD-0 on the sparsified detector model and a two stage BP+BP+OSD-0}\label{app:BPBPOSD}
As stated in the main text, a natural question arising from the sparsification technique we propose is how much information is lost when doing so. In fact, this technique can also understood as compressing the detector error model. In principle, the rank of the matrices defining both models should be the same by construction, so all possible faults and detectors can appear in both models. Anyway, in order to understand this, we conduct the following simulations for the bivariate bicycle codes in \cite{BBcodes} and compare those to the standard BP+OSD-0 (with $10,000$ maximum minimum-sum BP iterations) running over the full DEM:

\begin{itemize}
    \item BP+OSD-0 fully over the sparsified detector model. Here, we are oblivious to the fact that there is a DEM that has a richer information about how the faults behave at the circuit-level and directly attempt decoding over the sparsified detector model. We just map the priors using the transfer rule discussed in the text. The maximum number of minmum-sum BP iterations is set to $10,000$.
    \item Leveraging the two-stage BP proposed, resulting in a BP+BP+OSD-0 decoder. We set up the BP+BP decoder discussed in the main text running $1,000$ and $9,000$ maximum minimum-sum BP iterations over the DEM and the sparsified detector model, respectively.
\end{itemize}

Figure \ref{fig:BPOSDsparse} shows the obtained results. For the first case, it can easily be observed that being completely oblivious to the fact that we are compressing a bigger model results in a significant degradation in logical error rate. Interestingly, the decoder would still work and shows logical error rate reduction with code distance, but the performance is not on par with that of decoding over the full DEM. Another interesting thing that we observe by doing this relates with the mean time required for decoding. Note that, even if the sparsified detector model is significantly smaller than the DEM (recall that their sizes are around $30\%$ of the full model), we observed very similar mean decoding times for both scenarios. We consider that this comes from the fact that even if the model is smaller, the solutions to the problem will involve more columns of the sparsified model, indicating that both BP and OSD-0 will require some more time, resulting in the same thing as running the decoder over a much bigger model.

However, the second scenario resulted to be much more interesting. In Figure \ref{fig:BPOSDsparse}b, we can see that the performance of the BP+BP+OSD-0 decoder matches the one of BP+OSD-0. This is done by running just $10\%$ of the BP iterations over the full DEM. Furthermore, we measure that the BP+BP+OSD-0 requires $3\times$ less average decoding time than its counterpart. This comes from the fact that even if the smaller model requires solutions with higher weight, the BP iterations over the DEM and the transference to the sparsified model bias the solutions enough for faster convergence. This information that we get from the BP iterations over the DEM and their appropriate transference to the sparsified model does also explain why this decoder is able to match the accuracy of the BP+OSD-0 decoder. To sum up, the sparsified detector model can get enough information if some BP rounds are run over the full DEM. We consider that many other decoding proposals can benefit of this fact to improve their speed or to operate properly over circuit noise if they showed promise over phenomenological noise models but failed to escalate further.\\

\begin{figure*}[!ht]
    \centering
    \includegraphics[width=0.19\linewidth]{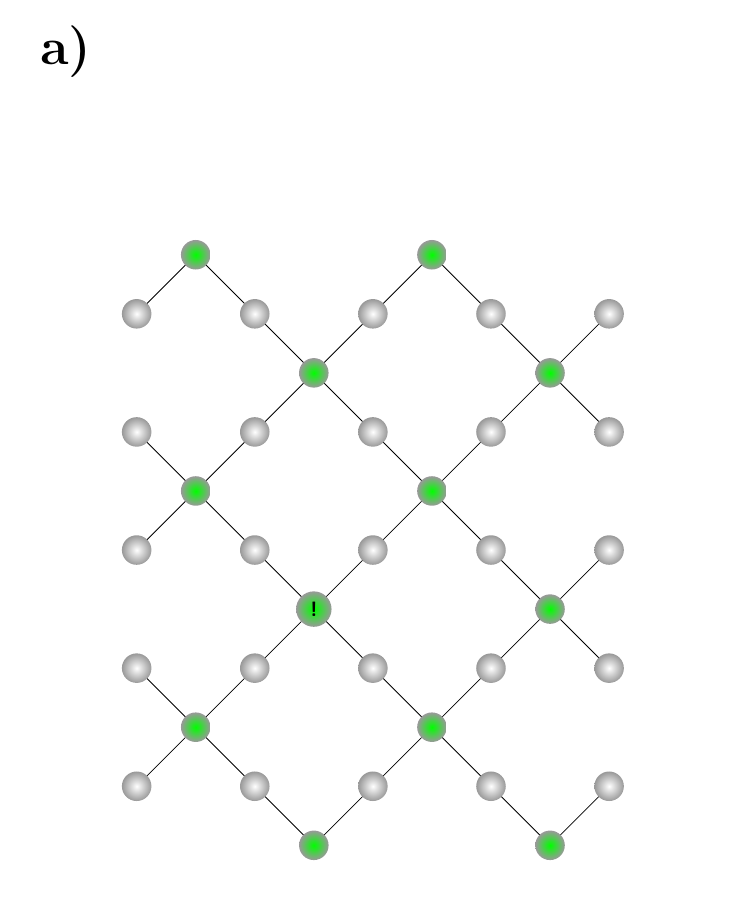}
    \includegraphics[width=0.19\linewidth]{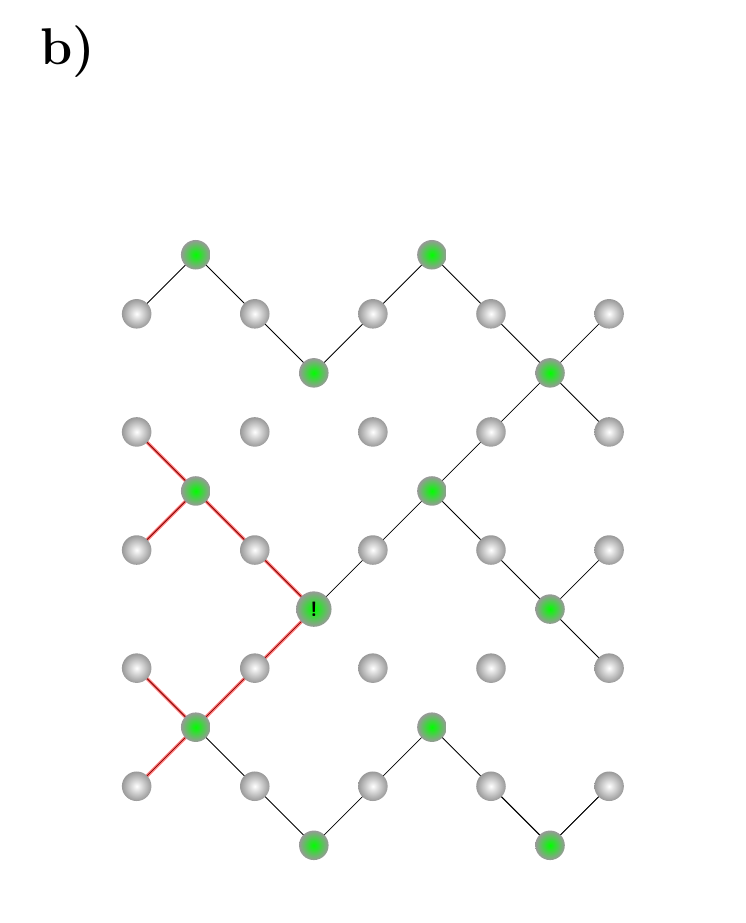}
    \includegraphics[width=0.19\linewidth]{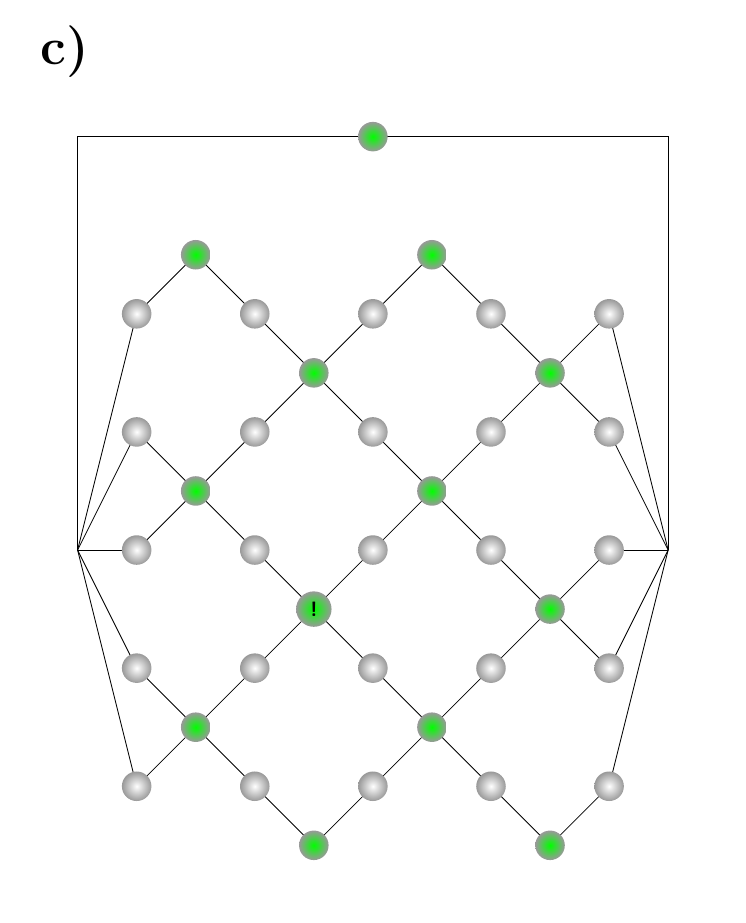}
    \includegraphics[width=0.19\linewidth]{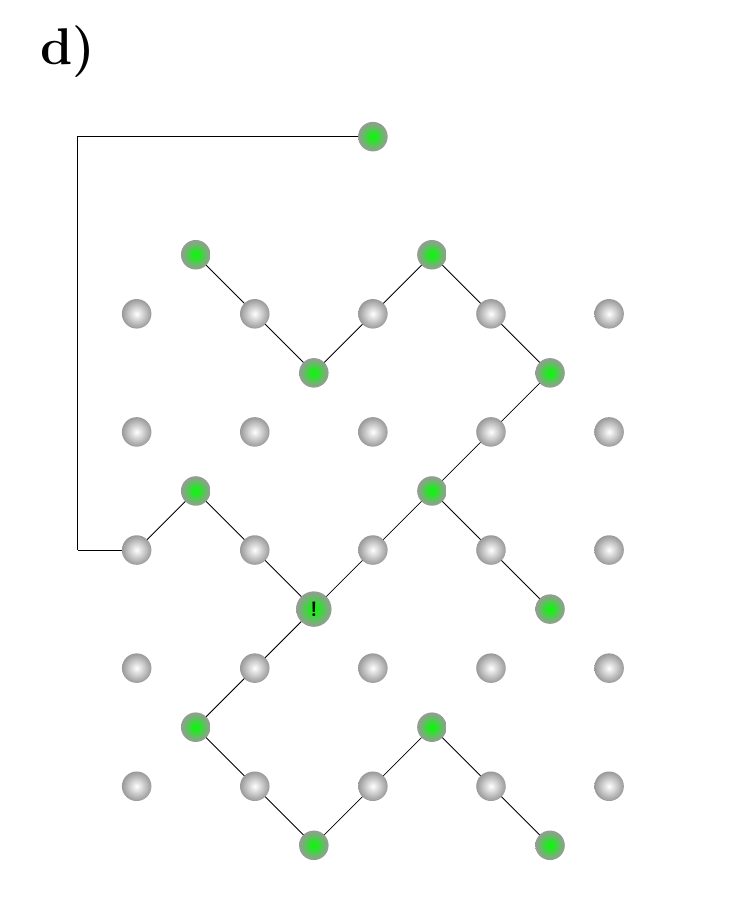}
    \includegraphics[width=0.19\linewidth]{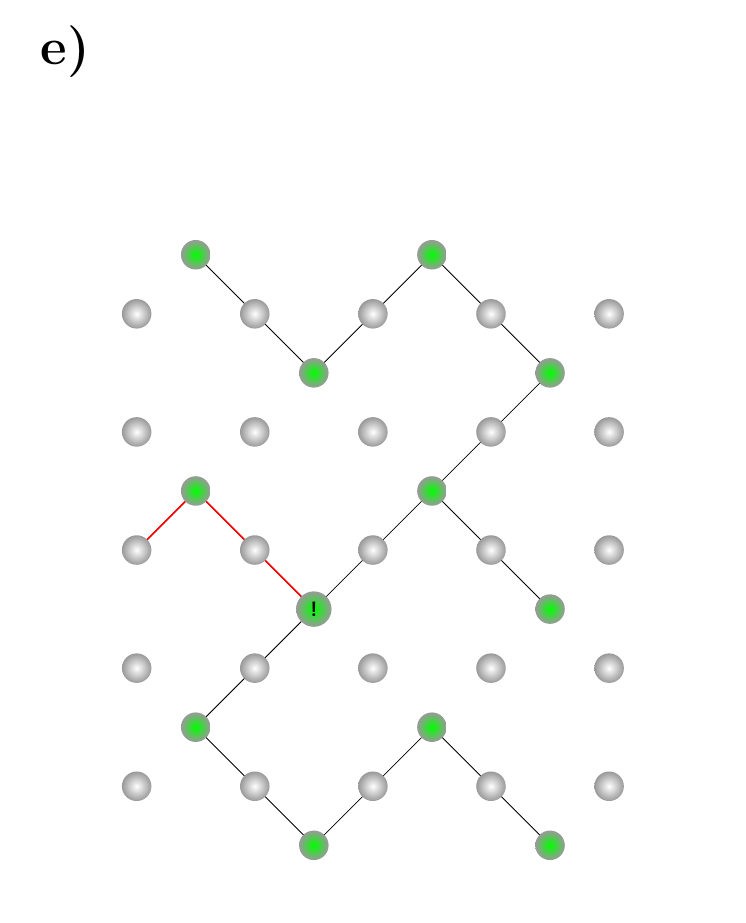}
    \caption{Representation of the process of finding an ordered Tanner forest for a specific syndrome in a surface code. In a) the syndrome is presented. In b) an ordered Tanner forest implementing Kruskal's algorithm directly in the Tanner graph can be seen. The red lines portray the variable nodes that will prevent the BP process from converging. In c) a virtual check presented on the top is introduced to the Tanner graph so as to run a Kruskal algorithm on it. In d) a Tanner forest is found which is then used in e) for finding an error recovery, also represented with a red line.}
    \label{fig:virtual_check}
\end{figure*}

\section{OTF in surface codes, considerations and virtual checks}\label{app:virtual}

The nature of surface codes prevents the OTF algorithm from being implemented directly into the parity check matrix. This is because variable nodes in the boundaries only have a single adjacent check. To the eyes of the modified Kruskal algorithm, variable nodes connected to a single check node cannot produce loops as it would require two adjacent checks with the same root. Consequently, all variable nodes adjacent to a single check will be accepted into any Tanner forest. This will produce a heavy detriment to the task of the second BP, as some sets of variable nodes can produce trapping sets that will prevent the second BP algorithm from preventing.

Fortunately, one can circumvent this phenomenon for surface codes by including a \textit{virtual check} to the Kruskal graph algorithm. This is a check only considered in Kruskal's algorithm section that is adjacent to all variable nodes which have parity check column weight 1. Including this check allows the Kruskal algorithm to avoid trapping sets consisting of boundary variable nodes. Figure \ref{fig:virtual_check} portrays the inclusion of the virtual check node when implementing Kruskal's algorithm for surface codes. For the simulations of surface codes, we have considered two different virtual checks, one for each different independent Tanner graph.

\end{appendix}
\end{document}